\documentclass{agujournal}		
\graphicspath{{Figures/}}

\journalname{JGR-Solid Earth}

\begin{document}

%
%


\title{Small-scale metal/silicate equilibration during core formation: the influence of stretching enhanced diffusion on mixing}

%
%

\authors{V. Lherm\affil{1}, R. Deguen\affil{1}}

\affiliation{1}{Universit\'e de Lyon, UCBL, ENSL, CNRS, Laboratoire de G\'eologie de Lyon : Terre, Plan\`etes, Environnement, 69622 Villeurbanne, France}



\correspondingauthor{Victor Lherm}{victor.lherm@ens-lyon.fr}




\begin{keypoints}
\item Stretching enhanced diffusion accelerates equilibration and mixing in stretched sheets and ligaments
\item Equilibration happens when the metal is stretched down to a scale (a modified Batchelor scale) smaller than $\sim 5$ cm
\item Impactors' cores are vigorously stretched and convoluted in a stirring regime
\end{keypoints}

%
%


\begin{abstract}
Geochemical data provide key information on the timing of accretion and on the prevailing physical conditions during core/mantle differentiation. However, their interpretation depends critically on the efficiency of metal/silicate chemical equilibration, which is poorly constrained.
Fluid dynamics experiments suggest that, before its fragmentation, a volume of liquid metal falling into a magma ocean undergoes a change of topology from a compact volume of metal toward a collection of sheets and ligaments.
We investigate here to what extent the vigorous stretching of the metal phase by the turbulent flow can increase the equilibration efficiency through what is known as stretching enhanced diffusion. We obtain scaling laws giving the equilibration times of sheets and ligaments as functions of a P\'eclet number based on the stretching rate. 
At large P\'eclet, stretching drastically decreases the equilibration time, which in this limit depends only weakly  on the diffusivity.
We also perform 2D numerical simulations of the evolution of a volume of metal falling into a magma ocean, from which we identify several equilibration regimes depending on the values of the P\'eclet (Pe), Reynolds (Re), and Bond (Bo) numbers. At large Pe, Re and Bo, the metal phase is vigorously stretched and convoluted in what we call a stirring regime. The equilibration time is found to be independent of viscosity and surface tension and depends weakly on diffusivity. Equilibration is controlled by an efficient thermochemical stretching enhanced diffusion mechanism developing from the mean flow and entraining the surrounding silicate phase.
\end{abstract}

%
%

%


%
%
%
%

\section{Introduction}
\label{sec:introduction}

The initial conditions of the Earth and other terrestrial planets of the Solar System are inherited from their concomitant accretion and differentiation.
Terrestrial planets differentiate into an iron core and a silicate mantle, inducing chemical fractionation and heat partitioning between the metal phase migrating toward the core and the surrounding silicate phase \citep{rubie_2015}.
The partitioning of heat provided by the impacts and the reduction of the potential gravitational energy sets the initial conditions of the temperature contrast between the core and the mantle \citep{rubie_2015}. This is crucial for the early thermal and magnetic evolution of planetary bodies \citep{monteux_2011,williams_2004}, and the formation and evolution of primitives magma oceans \citep{sun_2018}, with in particular the possibility of forming a basal magma ocean \citep{labrosse_2007}.
The partitioning of chemical elements has also profound geodynamical implications. 
For example, 
the identity and abundance of light and radioactive elements in the core \citep{badro_2015,corgne_2007}, which are key parameters for the dynamics and evolution of the core, depend on ($P$,$T$,$fO_2$) conditions at which metal and silicate have equilibrated.

Geochemical data such as geochronometers, estimates of the mantle and core composition, meteorite geochemistry, and partition coefficients \citep{li_1996,righter_2011,siebert_2011} constitute an inverse problem for the timing of accretion, the ($P$,$T$,$fO_2$) segregation conditions and the degree of chemical equilibration of the impactors' cores with the mantle of the Earth. This inverse problem is under-determined since there are more degree of freedom than constraints.
The interpretation of core formation chronometers well illustrates this issue since there is a strong trade-off between the estimated accretion time and the equilibration efficiency: the Hf-W chronometer predicts an accretion time around 30 Myr if perfect equilibration is assumed \citep{kleine_2002}, but that time is significantly increased in case of imperfect equilibration \citep{rudge_2010}.

The interpretation of geochemical data crucially depends on the equilibration efficiency of the segregation mechanisms. The scenario typically used in geochemical core formation models involves the impactor's core equilibration in a deep magma ocean (\textit{cf.} table \ref{tab:parameters_magma} for typical physical parameters) \citep{rubie_2015}.
The metal phase is assumed to fragment into centimetric drops, before separating quickly from the silicates as an \textit{iron rain} and sedimenting to the base of the magma ocean \citep{ichikawa_2010,karato_1997,rubie_2003,stevenson_1990}. The centimeter scale corresponds to the size at which surface tension prevents further fragmentation of the drops \citep{rubie_2003}. It is small enough to ensure efficient chemical equilibration between the droplets and the surrounding magma ocean \citep{rubie_2003}. Iron finally migrates through the solid part of the mantle toward the forming core by diapirism, initiated by Rayleigh-Taylor instabilities within the liquid pond \citep{karato_1997,monteux_2009,samuel_2010,stevenson_1990}, diking \citep{stevenson_2003} or percolation \citep{stevenson_1990}. No further equilibration is expected at this stage owing to the solid state of the underlying mantle and the large size of the diapirs.

\begin{table}[ht]
\caption{Typical parameters of a metallic impactor's core falling into a magma ocean. Modified after \citet{solomatov_2015} and \citet{deguen_2011}. Dimensionless numbers are calculated for a 100 km impactor's core.}
\centering
\begin{tabular}{rlll}
\hline
&\multicolumn{2}{c}{Magma ocean}\\
\hline
Magma ocean depth, $L$&\multicolumn{2}{c}{$10^6$}&m\\
Gravity, $g$&\multicolumn{2}{c}{10}&m.s$^{-2}$\\
\hline
&Silicate phase&Metal phase&\\
\hline
Density, $\rho_i$&$4\times 10^3$&$7.8\times 10^3$&kg.m$^{-3}$\\
Thermal expansion, $\alpha_i$&$5\times 10^{-5}$&$10^{-5}$&K$^{-1}$\\
Specific heat capacity, $c_{p_i}$&$10^{3}$&$5\times 10^2$&J.kg$^{-1}$.K$^{-1}$\\
Dynamic viscosity, $\eta_i$&$10^{-1}$&$10^{-2}$&Pa.s\\
Diffusion coefficient, $\kappa_i^c$&$10^{-9}$&$10^{-9}$&m$^{2}$.s$^{-1}$\\
Thermal conductivity, $\lambda_i$&4&$10^2$&W.m$^{-1}$.K$^{-1}$\\
Thermal diffusivity, $\kappa_i$&$10^{-6}$&$10^{-5}$&m$^{2}$.s$^{-1}$\\
Surface energy, $\gamma$&0.5&0.5&J.m$^{-2}$\\
\hline
Reynolds number, Re&$10^{12}$&$10^{14}$&-\\
Thermal P\'eclet number, $\mathrm{Pe}_{T}$&$10^{14}$&$10^{12}$&-\\
Compositional P\'eclet number, $\mathrm{Pe}_{C}$&$10^{17}$&$10^{17}$&-\\
Thermal Ohnesorge number, Oh$_T$&$10^{-12}$&$10^{-10}$&-\\
Compositional Ohnesorge number, Oh$_C$&$10^{-15}$&$10^{-15}$&-\\
Bond number, Bo&$10^{15}$&$10^{15}$&-\\
Weber number, We&$10^{15}$&$10^{15}$&-\\
\hline
\end{tabular}
\label{tab:parameters_magma}
\end{table}

\begin{figure}[ht!]
\centering
\includegraphics[width=\linewidth]{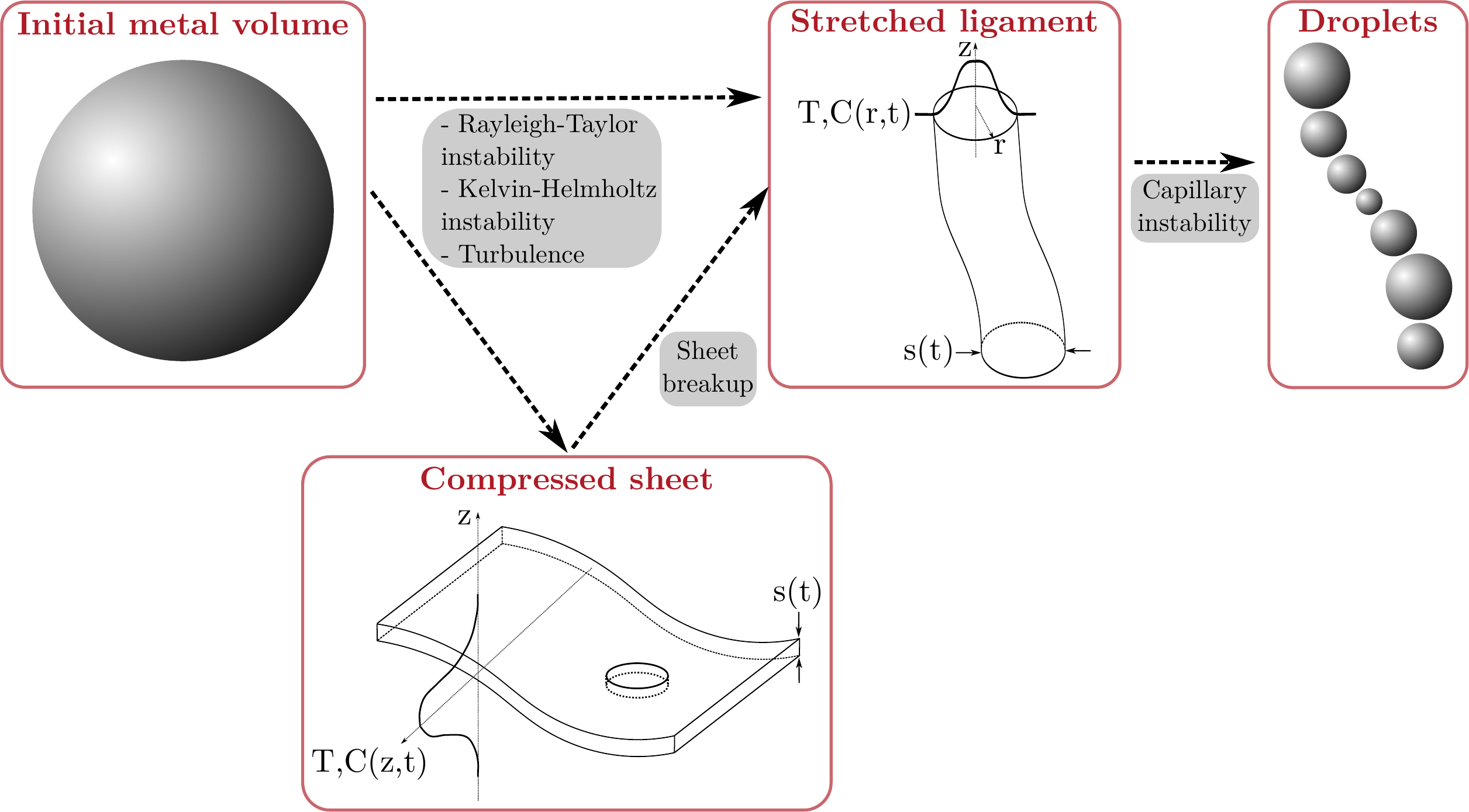}
\caption{A conceptual view of the fragmentation and equilibration of a metal volume falling into a magma ocean. Typical temperature/concentration profiles are represented for the sheet and the ligament.}
\label{fig:sed}
\end{figure}

The \textit{iron rain} scenario is based on the assumption of rapid fragmentation of the impactor cores into centimetric drops. Whether this is realistic remains an open question. Differentiation of terrestrial planets started early and most of the Earth was accreted from large differentiated planetesimals and embryos (100 to 1000 km) \citep{rubie_2015}. Whether the huge volumes of metal delivered by these impacts were indeed able to fragment into cm-scale drops is unclear \citep{dahl_2010,deguen_2014,kendall_2016,landeau_2014,wacheul_2014}. Liquid fragmentation is a well studied problem of fluid mechanics and the key mechanisms are reasonably well understood: though the details depends on the configuration of the flow, the route to fragmentation necessarily involves the formation of elongated liquid ligaments \citep[\textit{e.g.}][]{marmottant_2004a,villermaux_2004,villermaux_2007}, which is the only geometry unstable against the Rayleigh-Plateau capillary instability leading to fragmentation into drops.
Fragmentation must therefore be preceded by a change of topology of the metal phase, from a compact volume (the core of the impactor) toward metal ligaments.
Ligaments can be produced either directly by hydrodynamic instabilities or turbulence, or indirectly by the bursting of metal sheets, the liquid from the sheets then collecting into ligaments \citep[\textit{e.g.}][]{villermaux_2007}.
In the context of metal fragmentation in a magma ocean, laboratory experiments and numerical simulations suggest that the production of both metal ligaments and sheets can result from  a combination of Rayleigh-Taylor and shear instabilities, turbulent fluctuations, and interactions with the shear related to the crater opening and the following flow  \citep{deguen_2011,deguen_2014,kendall_2016,landeau_2014,wacheul_2014,wacheul_2018}. 
Once formed, ligaments can fragment into drops as a result  of the Rayleigh-Plateau capillary instability \citep{eggers_2008,marmottant_2004a,marmottant_2004}.
Figure \ref{fig:sed} illustrates the fragmentation sequence just described, which is to be considered as generic: if fragmentation of planetesimal's core indeed happens, it corresponds to the typical and necessary sequence of events required for the fragmentation of an initially spherical metallic impactor into stable droplets. 

In our conceptual fragmentation model (figure \ref{fig:sed}) and in laboratory experiments on which it is based \citep{deguen_2014,landeau_2014}, the liquid metal phase is vigorously stirred and stretched by the turbulent flow following the impact, before fragmentation actually happens. This suggests that mass and heat transfer between metal and silicates may be aided, or even controlled, by \textit{stretching enhanced diffusion}, a mechanism identified as crucial in the context of the mixing of a stirred diffusive heterogeneity \citep[\textit{e.g.}][]{coltice_2006,duplat_2008,kellogg_1987,olson_1984,ottino_1989,
ranz_1979,ricard_2015,venaille_2008,villermaux_2004}.
Stretching enhanced diffusion works as follows: take an initially compact patch of a diffusing tracer (temperature or concentration of solute) advected by a  flow. Unless the flow is uniform, the patch will be deformed according to the local strain tensor. In 2D, it will be stretched in one of the principal strain directions, and compressed in the other one (thus leading to the formation of 2D sheets). In 3D, it will be either stretched into one of the principal strain direction and compressed in the two others (thus leading to the formation of a filament), or compressed into one direction and elongated in the two others (leading to the formation of sheets). 
This stretching can drastically accelerate the homogenization of the tracer because the flow has the effect of increasing the exchange surface and maintaining strong concentration gradients in the compression direction(s), thus enabling efficient diffusive transport of the tracer. The effect is important if the stretching time (the inverse of the stretching rate) is smaller than the diffusion timescale based on the initial blob size.
For example, if an heterogeneity is stretched into a sheet of thickness $s(t)$  at a constant rate $\dot \epsilon = \mathrm{d}(\ln s)/\mathrm{d}t$, the homogenization timescale $t_{h}$ is given by 
\begin{equation}
t_{h} \sim \frac{1}{2\dot{\epsilon}}\mathrm{ln}\left[2 \frac{\dot{\epsilon}s_0^2}{K}+1\right],
\label{eq:t12_Ranz}
\end{equation}
where $s_0$ the initial thickness of the heterogeneity, and $K$ the tracer diffusivity (\textit{e.g.} \citet{ranz_1979}; see also \citet{kellogg_1987}).
If the stretching time $1/\dot \epsilon$ is large compared to the diffusion time $s_0^2/K$ (\textit{i.e.} if ${\dot{\epsilon}s_0^2}/{K}\ll 1$), a Taylor expansion of equation \ref{eq:t12_Ranz} shows that $t_{h}\to s_0^2/K$, which shows that in this limit the homogenization is controlled solely by diffusion.
However, if ${\dot{\epsilon}s_0^2}/{K}\gg 1$, equation \ref{eq:t12_Ranz} shows that $t_{h}$ depends predominantly on the stretching rate $\dot{\epsilon}$, and only weakly (logarithmically) on the tracer diffusivity $K$.
Stretching enhanced diffusion is a kinematic theory, and makes no assumption on the nature of the flow responsible for the stretching.
Though in deep Earth geodynamics this formalism has been used in the context of low-Reynolds, laminar flows (mixing of compositional heterogeneities in the mantle) \citep[\textit{e.g.}][]{coltice_2006,kellogg_1987,olson_1984}, it has been developed with turbulent flows in mind \citep{ranz_1979}, and has been used for example to characterize mixing in turbulent jets \citep[\textit{e.g.}][]{duplat_2008,villermaux_2004}.

Our goal is to test whether stretching enhanced diffusion can be an efficient mean of equilibrating (thermally and chemically) the metal phase of the impactors' cores with the immiscible molten silicates of the magma ocean, and to generalize its formalism (and equation \ref{eq:t12_Ranz} in particular) to heat and mass transfer between two stirred immiscible liquid phases.  
After presenting the governing equations and introducing a change of variable allowing to consider heat and mass transfer with the same set of equations (section \ref{sec:equations}), we build a regime diagram for the deformation of an initially round drop falling under its own weight  (section \ref{sec:path}).
Our numerical simulations and laboratory experiments \citep{deguen_2014,landeau_2014,wacheul_2014,wacheul_2018} show that a mass of metal falling into a magma ocean is most probably in a regime which we call \textit{stirring regime}, in which the metal is stretched into convoluted sheets and ligaments, before fragmenting into drops.
We then study in section \ref{sec:basic} the equilibration of what we consider to be the ``building blocks'' of the fragmentation and equilibration sequence, \textit{i.e.} sheets, ligaments and droplets.
In section \ref{sec:stirring}, we finally come back to the \textit{stirring regime} which we characterize in light of our results on the equilibration of stretched sheets.

\section{Equations and Numerical Model}
\label{sec:equations}

\subsection{Geometries}
\label{sec:equations-geometries}

\begin{figure}[ht!]
\centering
\includegraphics[width=1\linewidth]{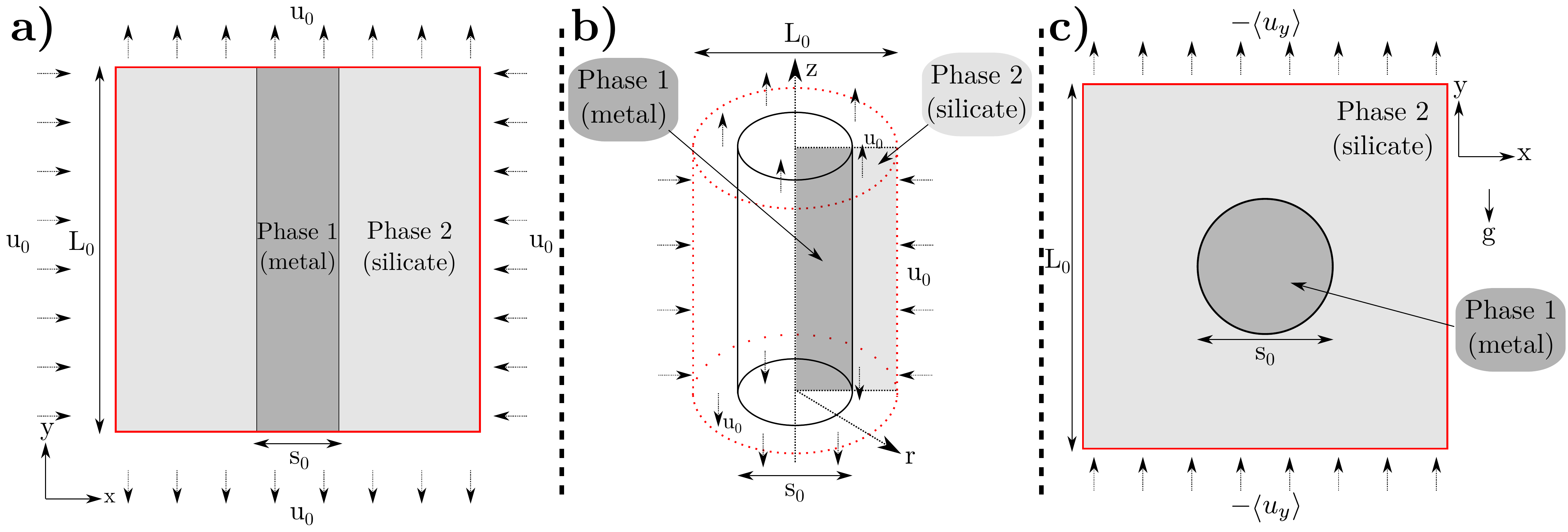}
\caption{Setup for the 2D sheet (a), ligament (b) and drop (c) calculations. Dashed arrows represent the imposed velocity field and the red lines indicate the numerical box boundary. (a): $L_0$ is the box size and $s_0$ the initial thickness of the sheet. (b): $L_0/2$ is the box size and $s_0$ the initial diameter of the ligament. The metal phase corresponding to the inner cylinder (black lines) is surrounded by the silicate phase, the shaded radial plan corresponds to the 2D numerical simulation area. (c): $L_0$ is the box size and $s_0$ the initial diameter of the drop. The imposed velocity field compensates the mean falling velocity $\langle u_y \rangle$.}
\label{fig:geometries}
\end{figure}

In sections \ref{sec:path}, \ref{sec:basic}, and \ref{sec:stirring}, we will carry out analytical and numerical calculations (\textit{cf.} table S1 for numerical parameters) on the temperature and concentration evolution of sheets, ligaments and drops.
In each geometry (figure \ref{fig:geometries}), we consider a two-phase flow involving a liquid silicate-like outer phase (phase 2) and a liquid metallic-like inner phase (phase 1).
Sheets and ligaments are stretched by a constant pure shear flow at a strain rate $\dot\epsilon$. We don't consider the effect of gravity. The two phases have different densities, viscosities, and diffusivities, both for analytical (sections \ref{sec:basic-sheet} and \ref{sec:basic-ligament-analytical} for sheets and ligaments, respectively) and numerical calculations (supplementary information and section \ref{sec:basic-ligament-tension} for sheet and ligaments, respectively). In numerical simulations, ligaments are free to fragment.
We also consider the case of a free-falling, initially round metal mass.
Numerical calculations on deformed drops assume identical viscosities and diffusivities between the two phases (sections \ref{sec:path}, \ref{sec:basic-droplet} and \ref{sec:stirring}).
Analytical calculations on non-deformed drops do take into account the conductivity and diffusivity contrasts (section \ref{sec:basic-droplet}).

\subsection{Governing equations, and equivalence between the heat and mass transfer problems}
\label{sec:equations-equivalence}

We study here the evolution of either temperature $T_i$ or concentration $C_i$ in a two-phase flow, where $i \in \lbrace 1,2 \rbrace$ refers to the metallic ($i=1$) or the silicate phase ($i=2$). 
In the Boussinesq approximation and in the absence of volumetric heat sources (radioactive heating or latent heat), the evolution of temperature and composition within both phases is governed by two transport equations of identical mathematical form. 
Mathematically speaking, the only difference between heat and mass transfer problems lies in the boundary conditions at the metal/silicates interface, which can be written
\begin{eqnarray}
\label{eq:boundaryT1}
T_1&=&T_2,\\
\label{eq:boundaryT2}
\lambda_1{\bf \nabla}T_1\cdot{\bf n}&=&\lambda_2{\bf \nabla}T_2\cdot{\bf n},
\end{eqnarray}
for temperature, 
and 
\begin{eqnarray}
\label{eq:boundaryC1}
D_{1/2}&=&\frac{C_1}{C_2},\\
\label{eq:boundaryC2}
\kappa_1^c{\bf \nabla}C_1\cdot{\bf n}&=&\kappa_2^c{\bf \nabla}C_2\cdot{\bf n},
\end{eqnarray}
for composition, where $\lambda_i$ is the thermal conductivity, $\kappa_i^c$ the mass diffusivity, $D_{1/2}$ the partition coefficient and ${\bf n}$ the unit vector normal to the interface pointing toward phase 2.
Equations \ref{eq:boundaryT2} and \ref{eq:boundaryC2} express the continuity of heat and mass flux at the interface, respectively.

Temperature is continuous across the interface, but concentration is not.
Yet the two problems can be made mathematically equivalent by introducing a new variable $\chi_i$ defined as either
\begin{equation}
\chi_i=T_i \quad\text{or}\quad \chi_i = \frac{C_i}{D_{1/2}^{\delta_{i1}}},
\end{equation}
where $\delta_{ij}$ is the Kronecker delta.
Rather than investigating separately heat and mass transfer, we will thus consider the scalar field $\chi_i$, which evolution is governed by the transport equation
\begin{equation}
\label{eq:AD}
\frac{\mathrm{D}\chi_i}{\mathrm{D}t}=K_i\nabla^2\chi_i,
\end{equation}
where $\mathrm{D\bullet}/\mathrm{D}t$ 
is the Lagrangian derivative, with boundary conditions
\begin{eqnarray}
\label{eq:general_BC1}
\chi_1&=&\chi_2,\\
\label{eq:general_BC2}
k_1{\bf \nabla}\chi_1\cdot{\bf n}&=&k_2{\bf \nabla}\chi_2\cdot{\bf n},
\end{eqnarray}
at the metal/silicate interface. $\chi_i$ has a diffusivity $K_i$ (\textit{i.e.} thermal diffusivity $\kappa_i$ or mass diffusivity $\kappa_i^c$) and a conductivity $k_i$ (equal to either $\lambda_i$ or $D_{1/2}^{\delta_{i1}} \kappa_i^c$). 
We will also make use of the ratio $q_i=k_i/K_i$, which is equal to either $\rho_i c_{p_i}$ or $D_{1/2}^{\delta_{i1}}$, where $\rho_i$ and $c_{p_i}$ are respectively density and specific heat capacity of the phase $i$.
The correspondence between the material properties of $\chi_i$ and their thermal and compositional counterparts are summarized in table \ref{tab:equivalence}.

\begin{table}[ht]
\caption{Summary of the correspondence between thermal and compositional cases. $\delta_{ij}$ is the Kronecker symbol. The general scalar field $\chi_i$ is associated with modified diffusivity $K_i$ and conductivity $k_i$ in each phase $i$. $q_i$ is also defined as the ratio between $k_i$ and $K_i$.}
\centering
\begin{tabular}{rcccc}
\hline
Thermal&$T_i$&$\kappa_i$&$\lambda_i$&$\rho_ic_{p_i}$\\
Compositional&$C_i/D_{1/2}^{\delta_{i1}}$&$\kappa_i^c$&$\kappa_i^c D_{1/2}^{\delta_{i1}}$&$D_{1/2}^{\delta_{i1}}$\\
\hline
General&$\chi_i$&$K_i$&$k_i$&$q_i$\\
\hline
\end{tabular}
\label{tab:equivalence}
\end{table}

In the following calculations, the velocity field will either be imposed, or obtained by solving the Navier-Stokes equation under the assumption of incompressibility,
\begin{eqnarray}
\label{eq:NS}
\rho_i\left(\frac{\partial{\bf u}}{\partial t}+{\bf u}\cdot\nabla{\bf u}\right)&=&-{\bf \nabla}P+\eta_i\nabla^2{\bf u}+\rho_i{\bf g},\\
\label{eq:imcomp}
\nabla\cdot{\bf u}&=&0,
\end{eqnarray}
where ${\bf u}$ is the flow velocity, $P$ is pressure, $\eta_i$ is viscosity, and $g$ is the acceleration of gravity.
The non-linear term on the left-hand side of equation \ref{eq:NS} is the source of turbulence. It is included in our numerical simulations, but since our simulations are 2D, we will not be able to reach fully developed turbulent regimes.
At the metal/silicates interface, the velocity field and tangential stress are continuous, while the normal stress is discontinuous due to interfacial tension,
\begin{equation}
\left[{\bf \sigma_1}-{\bf \sigma_2}\right]\cdot{\bf n}=\gamma\, ({\bf\nabla}\cdot{\bf n})\,{\bf n},
\label{eq:tension}
\end{equation}
where ${\bf \sigma_{i}}$ is the stress tensor, ${\bf\nabla}\cdot{\bf n}$ is the local curvature, and $\gamma$ is interfacial tension.

\subsection{Dimensionless numbers}
\label{sec:equations-dimensionless}

From these equations, we define a diffusion time $t_\kappa=s_0^2/K_{1}$, an advection time related to stretching $t_{\dot{\epsilon}}=1/\dot{\epsilon}$, a free-fall time related to gravity $t_g=\sqrt{(\rho_{1} /\Delta\rho) s_0/g}$, a viscous time $t_\nu=s_0^2/\nu_{1}$ and a capillary time $t_\gamma=\sqrt{\rho_{1} s_0^3/\gamma}$, with $s_0$ the typical length (initial thickness of the sheet, or diameter of the ligament or drop), $K_{1}$ the thermal or mass diffusivity of phase 1, $\dot{\epsilon}$ the stretching rate, $\rho_1$ the density of phase 1, $\Delta\rho$ the density difference, and $\nu_1$ the kinematic viscosity of phase 1.

Based on these timescales and the remaining parameters of the set of equations, we build two sets of dimensionless numbers depending on the geometry of the flow we consider.
In the case of a stretched sheet or ligament in which we ignore the effect of gravity, using an advection timescale equal to the stretching timescale allows to build the following set of independent dimensionless parameters:
\begin{equation}
\mathrm{Re}_1=\frac{s_0^2\dot{\epsilon}}{\nu_1},\
\mathrm{Pe}_1=\frac{s_0^2\dot{\epsilon}}{K_1},\ 
\mathrm{Oh}_{\kappa_1}=\frac{K_1\sqrt{\rho_1}}{\sqrt{\gamma s_0}},\
\frac{K_1}{K_2}, \ 
\frac{k_1}{k_2}, \ 
\frac{\rho_1}{\rho_2},\
\frac{\nu_1}{\nu_2},
\label{eq:numbers-stretching}
\end{equation}
where Re$_1$ is the Reynolds number, Pe$_1$ the P\'eclet number, and Oh$_{\kappa_1}$ the thermal/compositional Ohnesorge number. 
We will also make use of the Weber number defined as    
\begin{equation}
\mathrm{We}_1=\frac{\rho_1 s_0^3\dot{\epsilon}^2}{\gamma},
\label{eq:weber-def}
\end{equation}
which is equal to $\left( \mathrm{Oh}_{\kappa_1} \mathrm{Pe}_{1}\right)^{2}$.
In the case of a free-falling mass of metal, we use an advection timescale equal to the free-fall time and the problem is now described by the following set of independent dimensionless parameters:
\begin{equation}
\mathrm{Re}_1=\sqrt{\frac{\Delta\rho}{\rho_1}\frac{g s_0^3}{\nu_1^2}},\
\mathrm{Pe}_1=\sqrt{\frac{\Delta\rho}{\rho_1}\frac{g s_0^3}{K_1^2}},\ 
\mathrm{Bo}=\frac{\Delta\rho g s_0^2}{\gamma},\
\frac{K_1}{K_2}, \ 
\frac{k_1}{k_2}, \ 
\frac{\rho_1}{\rho_2},\
\frac{\nu_1}{\nu_2}.
\label{eq:numbers-free-fall}
\end{equation}
The Ohnesorge number can be obtained from Bo and Pe$_1$ as $\mathrm{Oh}_{\kappa_{1}}=\mathrm{Bo}^{1/2}/\mathrm{Pe}_{1}$. 

The Reynolds number defined either as $\mathrm{Re}_{1}=t_\nu /t_{\dot{\epsilon}}$ or $\mathrm{Re}_{1}=t_\nu /t_g$ compares inertia to viscous forces.
The P\'eclet number $\mathrm{Pe}_{1}=t_\kappa /t_{\dot{\epsilon}}$ or $\mathrm{Pe}_{1}=t_\kappa /t_g$ accounts for the relative importance of advective transport compared to diffusive transport.
The thermal/compositional Ohnesorge number $\mathrm{Oh}_{\kappa}=t_\gamma /t_\kappa$ compares the influence of surface tension to thermal or mass diffusion.
The number $\mathrm{Oh}_{\kappa_{1}}$ thus defined has the same form as the classical Ohnesorge number ($\nu_1\sqrt{\rho_1}/\sqrt{\gamma s_{0}}$), except here the $t_{\gamma}$ is compared to the scalar diffusion time rather than to the momentum diffusion time $t_{\nu}$. 
The Weber number $\mathrm{We}=(t_\gamma /t_{\dot{\epsilon}})^2$ accounts for the relative importance of fluid's inertia compared to its interfacial tension, and the Bond number $\mathrm{Bo}=(t_\gamma /t_g)^2$ for the relative importance of gravitational forces compared to tension forces.

\subsection{Numerical simulations with \textit{Basilisk}}
\label{sec:equations-basilisk}

The equations introduced in section \ref{sec:equations-equivalence} are solved in 2D with direct numerical simulations using the free software \textit{Basilisk} (see basilisk.fr and \citet{popinet_2009}). This partial differential equations solver uses an adaptive Cartesian mesh, making simulations possible up to Reynolds number of $10^4$. Mesh refinement is performed when the discretization error of a selected field (\textit{e.g.} temperature, concentration, velocity) is larger than an arbitrary value. It allows a good spatial resolution when and where it is needed, particularly in the vicinity of the stretched metal phase. On the contrary, spatial resolution is reduced far from the metal-silicate interface, where velocity variations are spatially smoother. 

The incompressible Navier-Stokes equations (eq. \ref{eq:NS} and \ref{eq:imcomp}) are solved with a Courant-Friedrichs-Lewy (CFL) condition limited time step, using a Bell-Collela-Glaz (BCG) advection scheme and a multigrid Poisson-Helmholtz solver for the viscous term. The BCG solver uses a second order upwind scheme.
The diffusion term of the advection-diffusion equation (eq. \ref{eq:AD}) is solved using a time implicit backward Euler discretization. The non-linear advection term of the equation is solved with the BCG scheme.

The two-phase flow is implemented using a level-set method. Each phase is characterized by its physical properties (\textit{e.g.} diffusivity, viscosity, etc) and the initial value of the scalar and vector fields (\textit{e.g.} temperature, concentration, velocity). The interface between the two phases is described by a level set function $\phi$ determining its shape.
The time evolution of the interface between the phases is computed with a Volume-Of-Fluid (VOF) advection scheme. First, the interface is reconstructed defining lines (2D) in each cell corresponding to the interface. Then, these lines and planes are moved with a geometrical flux computation. Surface tension (eq. \ref{eq:tension}) is finally implemented by computing the curvature of the interface.

\section{The Path Toward Equilibration: Regime Diagram and Timescales}
\label{sec:path}

\subsection{Dynamical shape regimes}
\label{sec:path_regime}

\begin{figure}[ht!]
\centering
\includegraphics[width=0.9\linewidth]{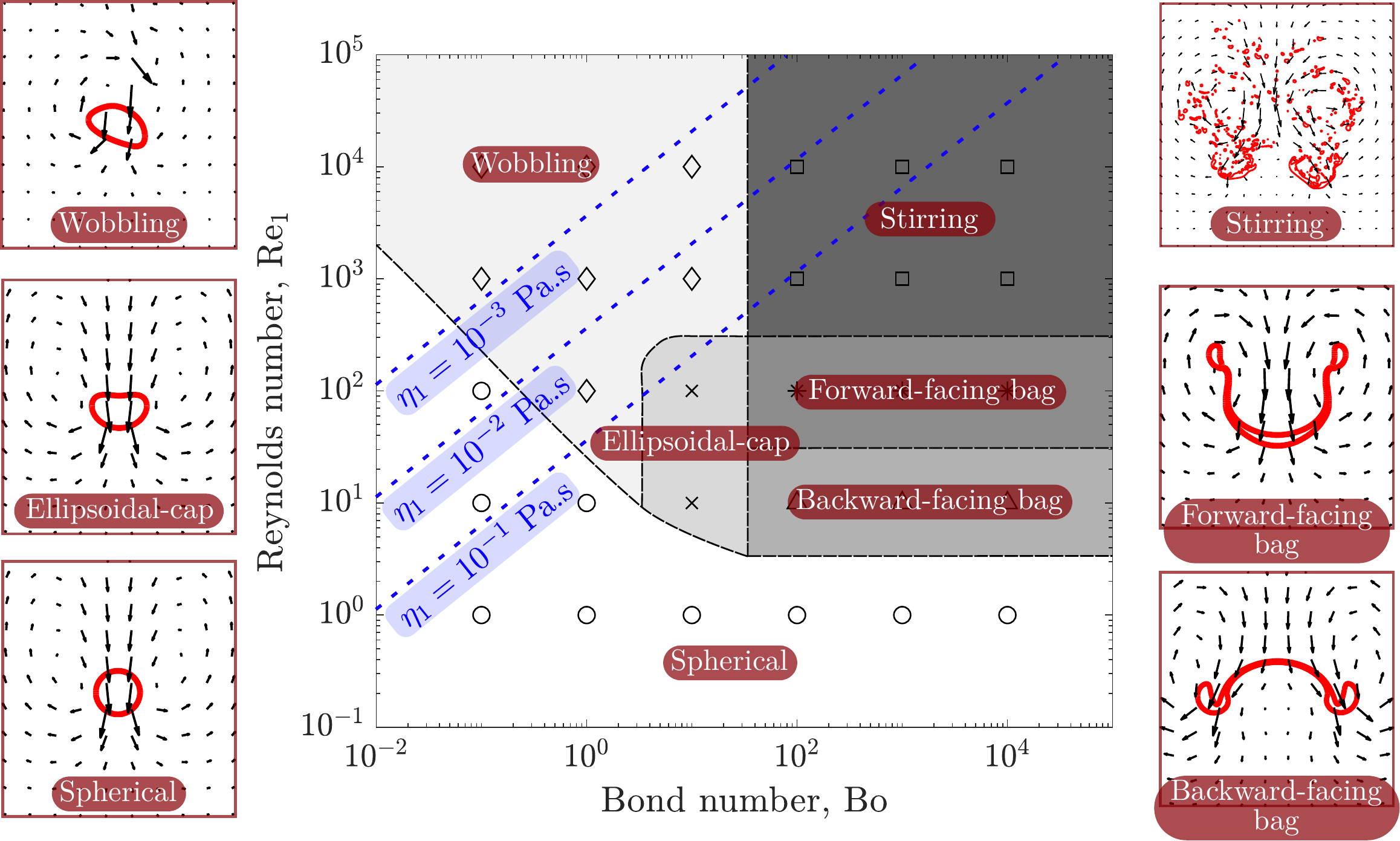}
\caption{Regime diagram of an initially round metal drop falling in a magma ocean, as a function of Re$_1$ and Bo. Markers correspond to numerical simulations.
The dashed blue lines correspond to the relationship between Re$_1$ and Bo (eq. \ref{eq:Re=f(Bo)}), for viscosities of the metal $\eta_1$ between $10^{-3}$ and $10^{-1}$ Pa.s. In the regime snapshots, the red lines correspond to the interface between metal and silicates and black vectors to the velocity field in a frame moving with the center of mass of the metal phase.}
\label{fig:drop_regime}
\end{figure}

We start here by building a regime diagram (figure \ref{fig:drop_regime}) for the deformation of an initially round metal volume falling from rest into molten silicates, from 2D numerical simulations.
We choose to construct a regime diagram as a function of the Reynolds number in the inner phase Re$_{1}$ (eq. \ref{eq:numbers-free-fall}, build with the inertial terminal velocity scaling) and of the Bond number Bo (eq. \ref{eq:numbers-free-fall}), which measures the relative importance of buoyancy and surface tension. Since we use $\nu_1=\nu_2$ in these simulations, $\mathrm{Re_2}=\sqrt{\rho_1/\rho_2}~\mathrm{Re_1}$ and the regime diagram is thus a function of $\mathrm{Re_2}$ divided by 1.4. Besides, viscosities in the metal and the magma ocean also vary, with a potential influence on the regime diagram \citep{wacheul_2018}.
Since in the case of a free-falling drop buoyancy is balanced by the largest of inertia and viscous forces, a small value of Bo implies that surface tension always dominates over the other forces, irrespectively of Re$_{1}$.
Conversely, a large value of Bo implies that surface tension is small compared to either viscous forces (low Re$_{1}$) or inertia (high Re$_1$).

Though based on 2D simulations, our results are broadly consistent with published results from  experiments and numerical simulations (in axisymmetric configurations)  \citep[\textit{e.g.}][]{clift_1978,han_1999,landeau_2014}.

At low Re$_1$, the drop remains undeformed, irrespectively of the value of Bo. 
The flow around the drop is laminar and stationary, and the drop develops an internal circulation. 
Keeping Bo small and increasing Re$_{1}$, the velocity field eventually becomes time-dependent, with the magnitude of the velocity fluctuations increasing with Re$_{1}$. 
The velocity fluctuations do deform the drop surface and modify its trajectory, leading to a \textit{wobbling regime} \citep{clift_1978}, but surface tension still remains large compared to the time-dependent inertial stresses, which prevents any significant stretching of the metal and limits the amount of deformation. 
At moderate values of Bo ($\sim10$) and Re$_{1}$ ($\sim10-10^{2}$), we identify an \textit{ellipsoidal-cap regime} \citep{clift_1978} similar to the spherical regime except for the flattened shape of the droplet.

Deformation of the drop is significant only at large values of Bo and moderate-to-large values of Re$_{1}$.
At $\mathrm{Bo}\gtrsim 10^{2}$, we find that the initially round drop deforms into a backward-facing bag shape for Re$_{1}$ on the order of 10 \citep{baumann_1992,clift_1978,han_1999,samuel_2012,thomson_1886}, and into a forward-facing bag shape (similar to the \textit{skirted} mode of \cite{clift_1978} and to the \textit{jellyfish} mode of \cite{landeau_2014}) for Re$_{1}$ on the order of $10^{2}$.
At even larger values of Re$_{1}$, we identify a regime which we will refer to as \textit{stirring regime} where the velocity field 
consists in two counter-rotating vortices, 
with smaller scale velocity fluctuations super-imposed. The metal phase is vigorously stretched and deformed into sheets. 
This regime is qualitatively similar to the \textit{immiscible turbulent thermal} regime observed experimentally by \citet{deguen_2014} and \citet{landeau_2014}, even though the 2D geometry of our simulations does not allow as much turbulence to develop.

From the dimensionless numbers definition of section \ref{sec:equations-dimensionless}, we write Re$_1$ as a function of Bo as
\begin{equation}
\label{eq:Re=f(Bo)}
\mathrm{Re_1}=\left(\frac{\gamma^3 \rho_1^2}{\Delta\rho g \eta_1^4}\right)^{1/4}\mathrm{Bo}^{3/4}.
\end{equation}
The factor in between parenthesis is the inverse of a Morton number \citep[\textit{e.g.}][]{clift_1978}. It only involves material properties in addition to $g$, and typically varies by a factor of 10 mainly due to uncertainties on the viscosity $\eta_1$, which is relatively small compared to Re$_{1}$ and Bo varying by orders of magnitude depending on the size of the metal. Equation \ref{eq:Re=f(Bo)} (with uncertainties) thus define the part of the regime diagram which is the most relevant for metal segregation in a magma ocean. 
This is shown  in figure \ref{fig:drop_regime} (dashed blue lines) for parameters values given in table \ref{tab:parameters_magma} and a viscosity of the metal phase in the range $10^{-3}-10^{-1}$ Pa.s \citep{rubie_2003,rubie_2015}. 
If starting from a volume of metal on the large Re and Bo side of the regime diagram (i.e.  Re$_{1}\gtrsim 10^3$ and Bo$\gtrsim 10^2$, which corresponds to a diameter of $\gtrsim 5$ cm), the falling metal would be expected to be in the stirring regime (turbulent thermal). If fragmentation produces droplets at moderate Re$_{1}$ and Bo, these droplets will most likely be in either the wobbling or spherical regimes, or possibly in the ellipsoidal-cap regime.

\subsection{Equilibration times}
\label{sec:path_equilibration}

\begin{figure}[ht!]
\centering
\includegraphics[width=1\linewidth]{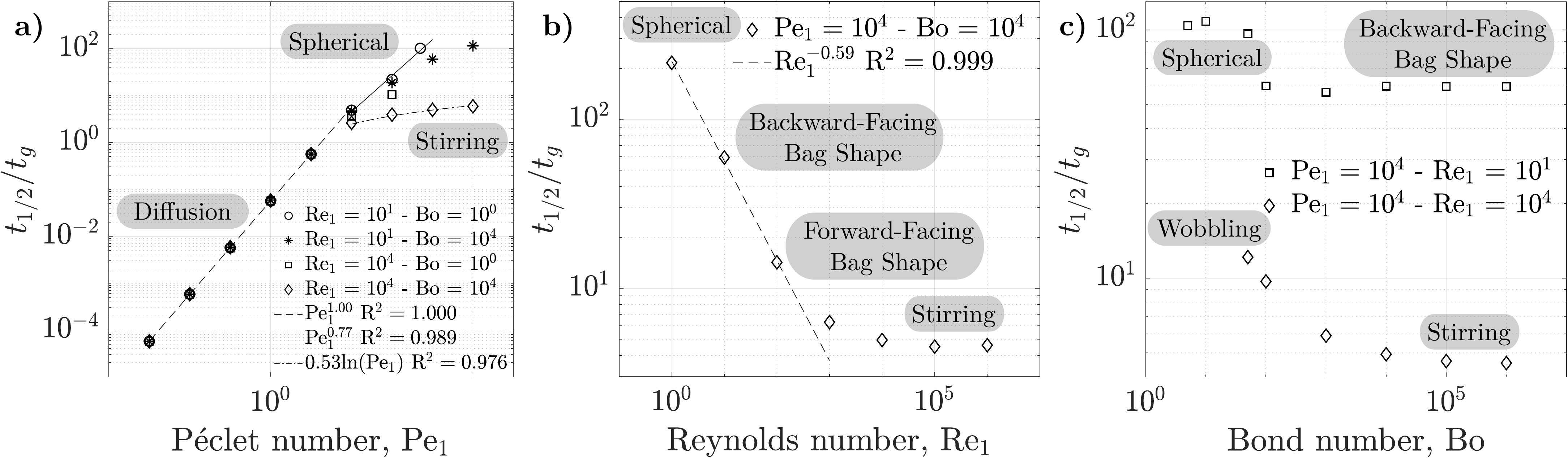} 
\caption{Equilibration time $t_{1/2}$ normalized by $t_{g}=\sqrt{(\rho_1/\Delta\rho)s_0/g}$ as a function of the P\'eclet number in the inner phase $\mathrm{Pe_1}$ (a), the Reynolds number in the inner phase $\mathrm{Re_1}$ (b), and the Bond number Bo (c). $t_{1/2}$ is the time at which $\langle \chi_1 \rangle$ is half its initial value. The lines correspond to calculated scaling law.}
\label{fig:drop_t12}
\end{figure}

We now focus on the thermochemical equilibration efficiency, which we characterize with an equilibration time $t_{1/2}$ corresponding to the time at which the mean scalar field in the metal phase $\langle \chi_1 \rangle$ is half its initial value.
Figure \ref{fig:drop_t12}a shows the equilibration time $t_{1/2}$ normalized by the free-fall time $t_g$ obtained from numerical simulations as a function of the P\'eclet number in the inner phase $\mathrm{Pe_1}$, 
for various values of Re$_1$ and Bo. 

At $\mathrm{Pe_1}\lesssim 10^2$, we find that $t_{1/2}/t_g \sim \mathrm{Pe_1}$, irrespectively of Re$_1$ and Bo. This implies $t_{1/2} \sim t_\kappa$, which means that equilibration is controlled by diffusion. 
Increasing Pe$_{1}$ above $\sim 10^{2}$ results in faster equilibration than in the diffusion regime.
At Pe$_{1}\gtrsim 10^{2}$ the equilibration time depends on the deformation regime of the drop, and hence on Re$_{1}$ and Bo, with the stirring regime being the most efficient at improving equilibration: we find $t_{1/2}/t_g \sim \mathrm{Pe_1}^{0.77}$ in the high Pe spherical regime (circles), while in the stirring regime, $t_{1/2}/t_g$ shows a weak dependency on Pe$_1$ (diamonds) with $t_{1/2}/t_g=0.53\,\mathrm{ln(Pe_1)}$ (the logarithmic Pe$_{1}$ dependency will be explained in section \ref{sec:stirring}).
The effect of the deformation regime can also be seen from figure \ref{fig:drop_t12}b which shows the effect on $t_{1/2}/t_g$ of increasing Re$_{1}$ while keeping Bo and Pe$_{1}$ constants at $10^{4}$.
At $\mathrm{Bo}=10^{4}$ the deformation regime changes from spherical to backward-facing bag shape, forward-facing bag shape, and finally stirring regime as Re$_{1}$ is increased. 
The equilibration time decreases significantly with Re$_{1}$ (as $t_{1/2}/t_g \sim \mathrm{Re_1}^{-0.59}$) while $\mathrm{Re_1 \lesssim 10^3}$, 
but $t_{1/2}/t_g$ seems to reach a plateau at Re$_{1}\gtrsim 10^{3}$ when entering the stirring regime. 
While higher Re$_{1}$ simulations would be needed to confirm this point, this suggests that the equilibration time may become independent of Re$_{1}$ (and hence on viscosity) in the stirring regime at high Re$_{1}$.
Figure \ref{fig:drop_t12}c shows $t_{1/2}/t_g$ as a function of the Bond number, for Pe$_{1}$ fixed at $10^{4}$ and Re$_{1}$ fixed at either 10 or $10^{4}$.
We find that $t_{1/2}/t_g$ decreases as Bo increases, irrespectively of Re$_1$, before reaching a plateau.
This is consistent with the fact that increasing Bo allows the metal phase to be more deformed, which helps equilibration.
This is particularly drastic in the stirring regime, in which the vigorous stretching and folding of the metal phase leads to very fast equilibration.

\section{Equilibration Processes in the ``Building Blocks'': Sheets, Ligaments and Droplets}
\label{sec:basic}

The regime diagram obtained in section \ref{sec:path}, and experiments \citep{deguen_2014,landeau_2014,wacheul_2014,wacheul_2018}, suggest that the most relevant dynamical regime at the beginning of the fragmentation sequence is the stirring regime, in which the metal phase is stretched into thin sheets (in our 2D simulations), and sheets and ligaments in 3D.
The metal phase will then eventually fragment into drops. 
We therefore study here heat and mass transfer in what we consider to be the three ``building blocks'' of the fragmentation and equilibration sequence: stretched sheets, ligaments, and drops.

\subsection{A stretched isolated sheet}
\label{sec:basic-sheet}

We consider here the homogenization of a scalar field $\chi(x,y,t)$ in and around a stretched isolated 2D sheet. 
Following \citet{ranz_1979}, the evolution of $\chi$ is described in a local Lagrangian frame $(x,y)$ which moves and rotates with the fluid such that the $y-$direction is always parallel to the stretching direction, and the $x-$direction perpendicular to it. 
In this frame of reference, the flow is a pure shear flow (or stagnation flow), and the velocity field is of the form $(u_x=-\dot{\epsilon}x,u_y=\dot{\epsilon}y)$, where $\dot{\epsilon}$ is the stretching rate. 
The thickness of the sheet $s(t)$ is given by $s(t)=s_0\mathrm{exp}\left(-\int_0^t\dot{\epsilon}(t')\mathrm{d}t'\right)$, and is assumed to be much smaller than the radius of curvature of the sheet.
Stretching will tend to align the direction of the gradient of $\chi$ with the compression direction, and we can therefore consider the sheet and the scalar field to be locally invariant in the stretching direction.
With this assumption and the velocity field given above, the scalar transport equation \ref{eq:AD} becomes
\begin{equation}
\frac{\partial \chi_i(x,t)}{\partial t}+\dot{\epsilon}x\frac{\partial \chi_i(x,t)}{\partial x}=K_i\frac{\partial^2\chi_i(x,t)}{\partial x^2}.
\label{eq:AD1D}
\end{equation}
We transform these equations into diffusion equations with the following change of variable \citep{ranz_1979}:
\begin{eqnarray}
\label{eq:adim_1}
\xi&=&\frac{x}{s(t)},\\
\label{eq:adim_2}
\tau&=&K_1\int_0^t\frac{\mathrm{d}t'}{s(t')^2}.
\end{eqnarray}
This amounts to normalize lengths by $s(t)$, which sets the thickness of the sheet to $\xi=1$, and time by a diffusion time scale based on the diffusivity inside the sheet (phase 1). 
With this change of variable, equation \ref{eq:AD1D} reduces to
\begin{equation}
\frac{\partial \chi_i(\xi,\tau)}{\partial \tau}=\frac{K_i}{K_1}\frac{\partial^2\chi_i(\xi,\tau)}{\partial \xi^2}.
\label{eq:D1D}
\end{equation}
The scalar field is initialized with a difference $\Delta\chi$ between the phases such as
\begin{eqnarray}
\chi_1(\vert\xi\vert \leq 1,\tau=0)&=&\Delta\chi,\\
\chi_2(\vert\xi\vert > 1,\tau=0)&=&0.
\end{eqnarray}
The boundary conditions are the continuity of the scalar field and its flux at the phase interface (equations \ref{eq:general_BC1} and \ref{eq:general_BC2}), and $\chi \rightarrow 0$ at infinity.

\begin{figure}[ht!]
\centering
\includegraphics[width=1\linewidth]{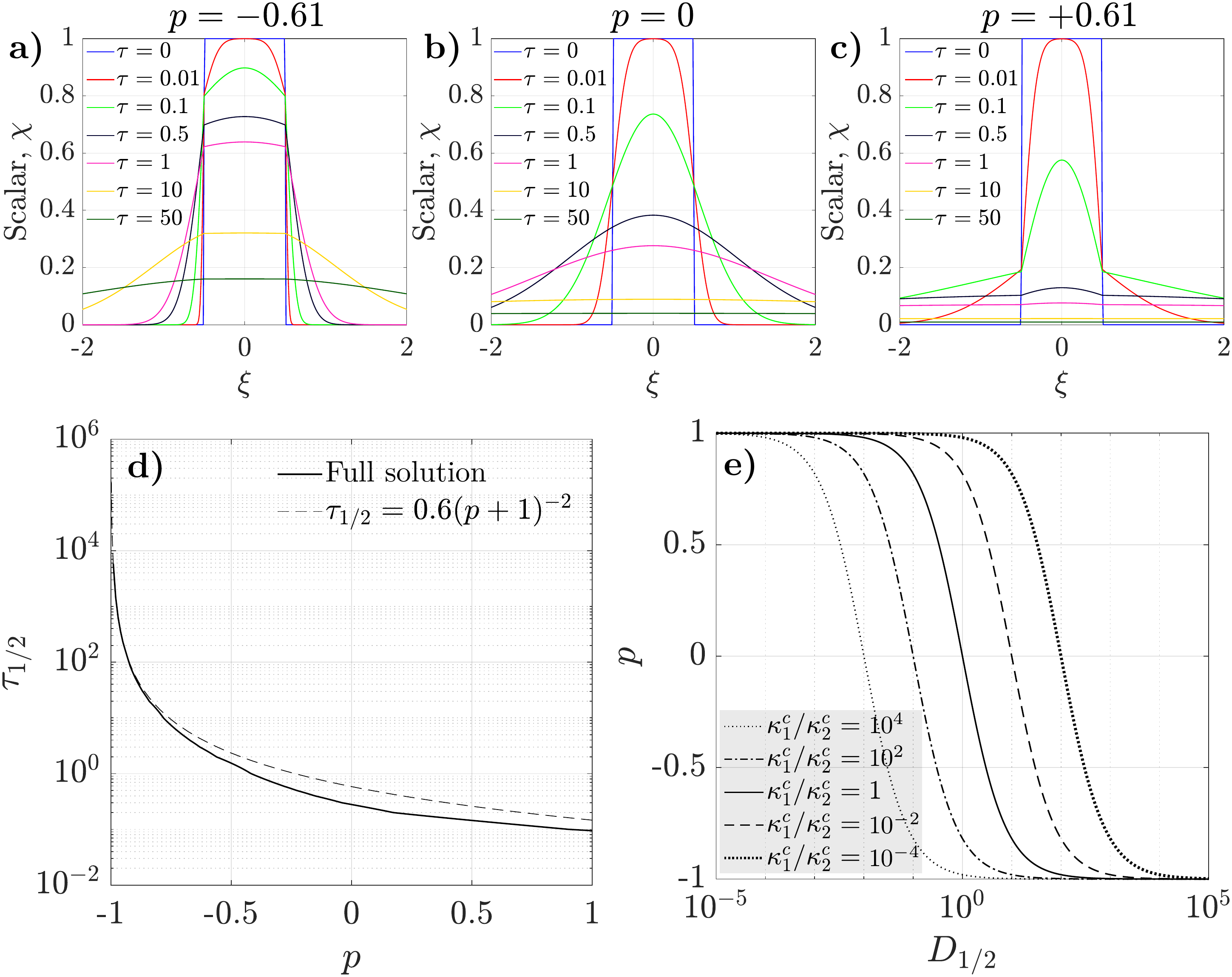}
\caption{Scalar profile in and around the sheet for $p=-0.61$ (a), $p=0$ \citep{jaupart_2010} (b) and $p=+0.61$ (c) as a function of $\tau$. $\tau$ and $\xi$ are respectively the normalized time and length (\textit{i.e.} equations \ref{eq:adim_1} and \ref{eq:adim_2}). (d): Evolution of $\tau_{1/2}$ as a function of $p$ (eq. \ref{eq:analytic_0}). The dashed line corresponds to the approximation of $\tau_{1/2}$ when $p \rightarrow -1$ (eq. \ref{eq:shape_f}). (e): Evolution of the parameter $p$ as a function of the partition coefficient $D_{1/2}$ for several values of the mass diffusivity ratio $\kappa_1^c/\kappa_2^c$ (from equation \ref{eq:p_general}).}
\label{fig:p_discussion}
\end{figure}

The full solution of this set of equations is given in appendix A, based on \citet{lovering_1936}.
In particular, the scalar in the middle of the sheet $\chi(\xi=0)$ evolves in time according to
\begin{eqnarray}
\label{eq:analytic_0}
\chi(\xi=0,\tau)=\Delta\chi(1+p)\sum_{n=0}^\infty (-p)^{n}~\mathrm{erf}\left(\frac{2n+1}{4\sqrt{\tau}}\right).
\end{eqnarray}
The parameter $p\in [-1,1]$ accounts for the diffusivity contrast in the two-phase flow and depends on the diffusivity ratio $k_1/k_2$ and on the ratio $q_1/q_2$ (see table \ref{tab:equivalence}) as
\begin{equation}
p=\frac{1-\sqrt{\frac{k_1}{k_2}\frac{q_1}{q_2}}}{1+\sqrt{\frac{k_1}{k_2}\frac{q_1}{q_2}}}.
\label{eq:p_general}
\end{equation}
Its value dramatically affects the scalar profiles (figure \ref{fig:p_discussion}b). If $p<0$ (figure \ref{fig:p_discussion}a), the diffusivity in the inner phase is larger than in the outer phase, leading to a flatter profile in the inner phase. On the contrary, if $p>0$ (figure \ref{fig:p_discussion}c), the diffusivity in the outer phase is larger than in the inner phase, leading to a flatter profile in the outer phase.

We compute from equation \ref{eq:analytic_0} the normalized equilibration time $\tau_{1/2}$ at which $\chi(\xi=0,\tau_{1/2})=\Delta\chi/2$. 
Equation \ref{eq:analytic_0} shows that $\tau_{1/2}$ is a function of $p$ only, and figure \ref{fig:p_discussion}d shows that $\tau_{1/2}$ decreases when $p$ increases. An approximate expression for $\tau_{1/2}$ can be found in the limit of $p$ close to -1, corresponding to $k_1 q_1 \gg k_2 q_2$,  as follows. In this limit, we expect equilibration to be limited by diffusion in the outer phase 2. The equilibration time $t_{1/2}$, corresponding to the normalized equilibration time $\tau_{1/2}$, should therefore be independent of $k_1$. According to equation \ref{eq:adim_2}, $t_{1/2}$ is a function of $\tau_{1/2}/K_1 = \tau_{1/2}\, q_1/k_1$, the form of which depends on the evolution of the thickness of the sheet. $\tau_{1/2}(p)$ must therefore be proportional to $k_1$. 
If $p \rightarrow -1$, $p$ simplifies as $p \sim -1+2\sqrt{\frac{k_2}{k_1}\frac{ q_2}{q_1}}$ from equation \ref{eq:p_general}. 
It thus satisfies the proportionality with $k_1$ only if
\begin{equation}
\label{eq:shape_f}
\tau_{1/2}(p) \sim \frac{1}{(p+1)^{2}},
\end{equation}
which in the limit $p\to -1$ tends toward $\sim \frac{1}{4}\frac{k_1}{k_2}\frac{q_1}{q_2}$.
Figure \ref{fig:p_discussion}d shows that the prediction of equation \ref{eq:shape_f} with a proportionality factor equal to 0.6 is indeed very close to the full solution of equation \ref{eq:analytic_0} when $p$ approaches -1. 
It is still reasonably accurate at higher values of $p$: it overestimates $\tau_{1/2}$ by a factor 2 at most over the all range of $p$.

In the heat transfer case, the value of $p$ in a magma ocean (table \ref{tab:parameters_magma}) is $p\sim-0.6$. In the composition case, $p=(1-\sqrt{\kappa_1^c/\kappa_2^c}D_{1/2})/(1+\sqrt{\kappa_1^c/\kappa_2^c}D_{1/2})$  depends on the partition coefficient and the mass diffusivity contrast between metal and silicates (figure \ref{fig:p_discussion}e). 
In particular, it can be seen that $p$ approaches $-1$  for siderophile elements (\textit{e.g.} W, Co, Cr, V, Ni), since $D_{1/2} \gg 1$ and $\kappa_1^c/\kappa_2^c$ is larger than unity \citep{oneill_1998}.

We now go back to the stretching of the metal sheet and calculate the equilibration time $t_{1/2}$ from equation \ref{eq:adim_2} taken at $\tau=\tau_{1/2}$. For that purpose, we assume that $\dot{\epsilon}$ is constant, as predicted in the case of homogeneous turbulence \citep{batchelor_1952}, which implies an exponential decrease of the layer thickness
\begin{equation}
s(t)=s_0\mathrm{exp}\left(-\dot{\epsilon}t\right).
\label{eq:thickness_evol}
\end{equation}
Using equation \ref{eq:adim_2}, we obtain
\begin{equation}
t_{1/2}=\frac{1}{2\dot{\epsilon}}\mathrm{ln}\left[2\mathrm{Pe_1}\tau_{1/2}(p)+1\right]=\frac{1}{2\dot{\epsilon}}\mathrm{ln}\left[\frac{2s_0^2\dot{\epsilon}}{K_1}\tau_{1/2}(p)+1\right],
\label{eq:t12_hypo0}
\end{equation}
which is a generalization of equation \ref{eq:t12_Ranz} to the homogenization of a scalar field between two liquid phases with different transport properties.
In the small P\'eclet limit, a Taylor expansion of equation \ref{eq:t12_hypo0} shows that $t_{1/2} \simeq \tau_{1/2}(p)\, s_{0}^{2}/K_{1}$, as expected. 
In the large P\'eclet limit when advection dominates over diffusion, \textit{i.e.} in the stretching enhanced diffusion regime, equation \ref{eq:t12_hypo0} simplifies to
\begin{equation}
t_{1/2}\sim\frac{1}{2\dot{\epsilon}}\mathrm{ln}\left[\mathrm{Pe_1}\tau_{1/2}(p)\right].
\label{eq:t12_hypo1}
\end{equation}
In the limit $p \rightarrow -1$ relevant to siderophile elements (figure \ref{fig:p_discussion}e), using equation \ref{eq:shape_f} gives
\begin{equation}
t_{1/2}\sim\frac{1}{2\dot{\epsilon}}\mathrm{ln}\left[\mathrm{Pe_1}\frac{1}{(p+1)^2}\right]\sim\frac{1}{2\dot{\epsilon}}\mathrm{ln}\left[\mathrm{Pe_2}\left(\frac{q_1}{q_2}\right)^2\right],
\label{eq:t12_hypo2}
\end{equation}
with $\mathrm{Pe}_2$ the P\'eclet number of the phase 2 defined in the same way as $\mathrm{Pe}_1$ (eq. \ref{eq:numbers-free-fall}).

\begin{figure}[ht!]
\centering
\includegraphics[width=1\linewidth]{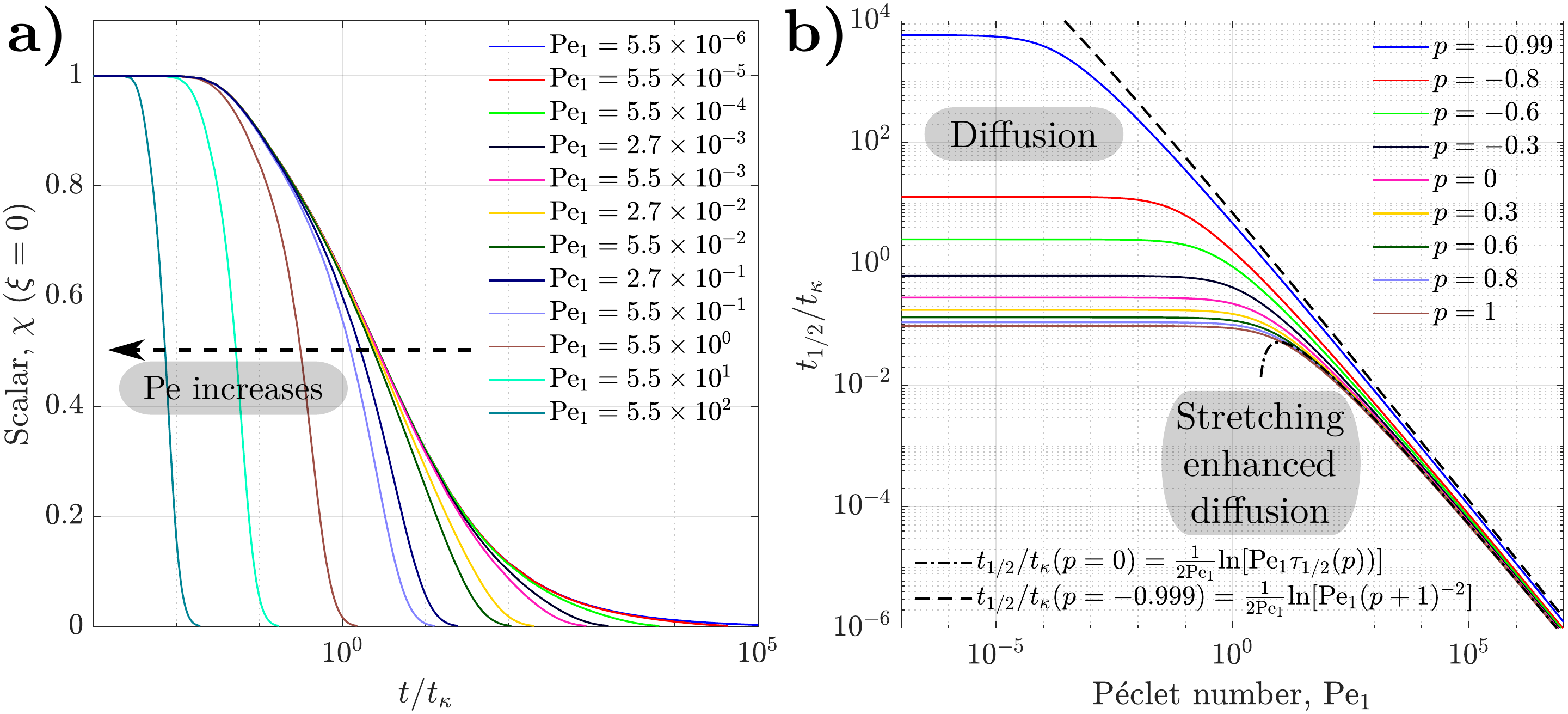}
\caption{(a): Value of $\chi_{1}$ in the middle of the sheet (calculated from equation \ref{eq:analytic_0} at $p=-0.61$) as a function of the time normalized by $t_{\kappa}=s_0^2/K_1$, for several values of the P\'eclet number in the inner phase $\mathrm{Pe_1}$.
(b): Equilibration timescale $t_{1/2}$ normalized by $t_{\kappa}=s_0^2/K_1$ as a function of the P\'eclet number in the inner phase $\mathrm{Pe_1}$, for several values of $p$ (from equation \ref{eq:t12_hypo0}).
The dash-dotted line corresponds to the large Pe approximation of equation \ref{eq:t12_hypo1}, calculated for $p=0$. The dashed line corresponds to the large Pe and $p \rightarrow -1$ approximation of equation \ref{eq:t12_hypo2}, calculated for $p=-0.99$.}
\label{fig:sheet_t12}
\end{figure}

Figure \ref{fig:sheet_t12}a shows the evolution of $\chi$ in the middle of the sheet as a function of the time normalized by $t_{\kappa}=s_0^2/K_1$, as given by equation \ref{eq:analytic_0}. If $\mathrm{Pe_1} \lesssim 1$, the equilibration time is nearly independent of $\mathrm{Pe_1}$, whereas if $\mathrm{Pe_1} \gtrsim 1$, $\chi(\xi=0)$ decreases over a timescale significantly reduced as $\mathrm{Pe_1}$ increases.
Figure \ref{fig:sheet_t12}b shows the equilibration time $t_{1/2}$ as a function of $\mathrm{Pe_1}$ according to equation \ref{eq:t12_hypo0}.
If $\mathrm{Pe_1}\ll 1$, $t_{1/2}/t_\kappa$ is independent of $\mathrm{Pe_1}$ and $t_{1/2}$ scales as the diffusion time $t_{\kappa}$, consistently with the low P\'eclet number approximation of equation \ref{eq:t12_hypo0}. We thus obtain the diffusive regime expected in that range of P\'eclet number (figure S1).
On the contrary, if $\mathrm{Pe_1}\gg 1$, $t_{1/2}/t_\kappa$ is well approximated by equation \ref{eq:t12_hypo1} with a weaker dependence on $p$ than in the diffusive regime (figure \ref{fig:sheet_t12}b, dash-dotted line).
If in addition $p \rightarrow -1$, $t_{1/2}/t_\kappa$ scales as equation \ref{eq:t12_hypo2} at large $\mathrm{Pe_1}$ (figure \ref{fig:sheet_t12}b, dashed line). $t_{1/2}$ depends predominantly on the stretching rate with a comparatively weak (logarithmic) dependency on the diffusivity.
If the stretching rate $\dot{\epsilon}$ increases, the equilibration time $t_{1/2}$ decreases dramatically (eq. \ref{eq:t12_hypo0} and figure \ref{fig:sheet_t12}b). In other words, if stretching intensifies, the equilibration efficiency between metal and silicates increases. In that case, we obtain a stretching enhanced diffusion regime (figure S2).
The critical $\mathrm{Pe_1}$ separating the two regimes depends on $p$. In particular, when $p\to -1$, \textit{i.e.} for large partition coefficient and diffusivity ratio $\kappa_1^c/\kappa_2^c$, the critical $\mathrm{Pe_1}$ is significantly decreased below 1.

Using equations \ref{eq:thickness_evol} and \ref{eq:t12_hypo1}, we can also calculate the width of the sheet $s_{1/2}$ at $t=t_{1/2}$.
In the $\mathrm{Pe_1} \gg 1$ limit, this gives
\begin{equation}
\label{eq:s12_sheet}
s_{1/2} \sim \frac{1}{\sqrt{\tau_{1/2}(p)}}\left(\frac{K_1}{\dot{\epsilon}}\right)^{1/2},
\end{equation}
where $(K_1/\dot{\epsilon})^{1/2}$ is a characteristic diffusion length, known  as the Batchelor length scale \citep{kellogg_1987}. $s_{1/2}$ corresponds to the length at which diffusion is no longer negligible, the sheet being sufficiently thin for diffusion to complete equilibration.
A 100 km metallic core falling at 100 m.s$^{-1}$ \citep{deguen_2011} produces a large scale stretching rate of $\mathrm{100~m.s^{-1}/100~km=10^{-3}~s^{-1}}$. Using a thermal diffusivity of $10^{-5}$ m$^2$.s$^{-1}$ (table \ref{tab:parameters_magma}), we obtain a thermal equilibration scale around 6 cm.
Concerning chemical equilibration, the $p \rightarrow -1$ limit of $\tau_{1/2}(p)$ (eq. \ref{eq:shape_f}) gives
\begin{equation}
\label{eq:s12_sheet_chemical}
s_{1/2} \sim \frac{2}{D_{1/2}}\left(\frac{\kappa_2^c}{\dot{\epsilon}}\right)^{1/2}.
\end{equation}
Using a mass diffusivity of $10^{-9}$ m$^2$.s$^{-1}$ and $D_{1/2}$ equal to 1, 10 and 100, the width of chemical equilibration is around 2, 0.2 and 0.02 mm, respectively.
Thermal equilibration may happen prior to fragmentation because the typical fragmentation scale corresponding to a capillary length is around 3 mm.
However, the chemical equilibration length is of the same order as the fragmentation scale, or even smaller regarding siderophile elements. It means that chemical equilibration may happen at the same time, or after, fragmentation of the metallic core occurs.

\subsection{A stretched isolated ligament}
\label{sec:basic-ligament}

\subsubsection{Analytical solution}
\label{sec:basic-ligament-analytical}

Using the same change of variable as for the sheet (equations \ref{eq:adim_1} and \ref{eq:adim_2}) allows to obtain an analytical solution of the axisymmetric advection-diffusion equation of the stretched ligament (figure \ref{fig:geometries}b). The scalar field and its flux are continuous across the phase interface.
From \citet{carslaw_1959}, p. 346, we obtain the solutions given in appendix B.

Using the same method as for the sheet (eq. \ref{eq:adim_2} and \ref{eq:thickness_evol}), we obtain a law for the equilibration time similar to equation \ref{eq:t12_hypo0}, except for $\tau_{1/2}$ which now depends independently on $K$ and $k$ (figure S8), leading to
\begin{equation}
t_{1/2}=\frac{1}{2\dot{\epsilon}}\mathrm{ln}\left[2\mathrm{Pe_1}\tau_{1/2}(K,k)+1\right].
\label{eq:t12_hypo0_ligament}
\end{equation}
We obtain a good agreement between the analytical solution from equation \ref{eq:t12_hypo0_ligament} (figure \ref{fig:ligament_t12}a, red circles) and numerical simulations (figure \ref{fig:ligament_t12}a, black circles).

As in section \ref{sec:basic-sheet}, we obtain the equilibration length, which can be written
\begin{equation}
\label{eq:s12_ligament}
s_{1/2} \sim \frac{1}{\sqrt{\tau_{1/2}(K,k)}}\left(\frac{K_1}{\dot{\epsilon}}\right)^{1/2}
\end{equation}
in the large P\'eclet limit.
For a 100 km metallic core falling at 100 m.s$^{-1}$ \citep{deguen_2011} and using typical parameters of table \ref{tab:parameters_magma}, we obtain $\tau_{1/2}(K,k) \sim 5.5$ and a thermal equilibration length around 4 cm. The chemical equilibration length is around 1.6, 0.5 and 0.1 mm for $D_{1/2}$ equal to 1, 10 and 100, respectively. As for the sheet in section \ref{sec:basic-sheet}, thermal equilibration may happen prior to fragmentation whereas chemical equilibration may happen during or after fragmentation.

\subsubsection{Effect of surface tension}
\label{sec:basic-ligament-tension}

Ligaments are unstable against the Rayleigh-Plateau capillary instability, and will eventually fragment into drops, possibly affecting equilibration.
Fragmentation (figure S5) results from the development of the capillary instability, which grows on a timescale on the order of $t_\gamma$ (capillary time based on the ligament initial diameter) in the absence of significant stretching, if $t_{\gamma} \ll t_{\dot\epsilon}$.
It is known that the capillary instability can be damped if the ligament is stretched \citep[\textit{e.g.}][]{eggers_2008,mikami_1975,taylor_1934,tomotika_1936}; stretching can delay fragmentation if $t_{\dot{\epsilon}} \ll t_\gamma$, \textit{i.e.} if $\mathrm{We_1}\gg 1$.

\begin{figure}[ht!]
\centering
\includegraphics[width=1\linewidth]{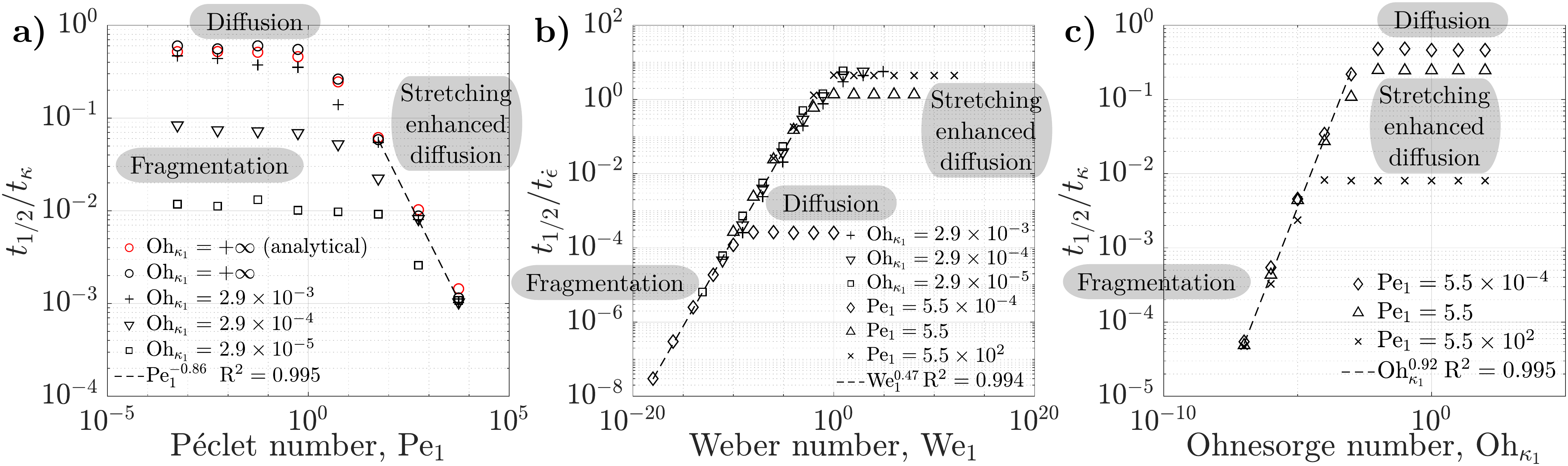}
\caption{Equilibration time $t_{1/2}$ normalized by $t_{\kappa}=s_0^2/K_1$ (a and c) or $t_{\dot{\epsilon}}=\dot{\epsilon}^{-1}$ (b) as a function of the P\'eclet number $\mathrm{Pe_1}$ (a), the Weber number $\mathrm{We_1}$ (b) and the thermal/compositional Ohnesorge number $\mathrm{Oh}_{\kappa_1}$ (c), in the inner phase.
Red circles corresponds to the analytical solution presented in equation \ref{eq:t12_hypo0_ligament}. The dashed line corresponds to the calculated power-law trend.}
\label{fig:ligament_t12}
\end{figure}

Figure \ref{fig:ligament_t12} shows the equilibration time $t_{1/2}$, defined here as the time at which the mean value of $\chi$ of the inner phase $\langle \chi_1\rangle$ is half its initial value, obtained from numerical calculations as a function of Pe$_{1}$, We$_{1}$, and Oh$_{\kappa_{1}}$.
Fragmentation happens after equilibration if $t_{\gamma}$ is large compared to the (no surface tension) equilibration time given by equation \ref{eq:t12_hypo0_ligament}.
This would be the case if $t_{\gamma}$ is large compared to either $t_{\kappa}  \tau_{1/2}$ (at small P\'eclet) or $t_{\dot{\epsilon}}\ln (\mathrm{Pe_{1}} \tau_{1/2})$ (at large P\'eclet), \textit{i.e.} if $\mathrm{Oh}_{\kappa_1}/\tau_{1/2}\gg 1$ or $\mathrm{We_1} \gg \left[ \ln (\mathrm{Pe_{1}} \tau_{1/2}) \right]^{2}$.
Fragmentation has no effect on equilibration in this limit, and the equilibration time obtained from the numerical calculations is consistent with the prediction of equation \ref{eq:t12_hypo0_ligament}. 
The flat slopes at large $\mathrm{We_1}$ (figure \ref{fig:ligament_t12}b) and $\mathrm{Oh_{\kappa_1}}$ (figure \ref{fig:ligament_t12}c) show that equilibration is indeed independent of surface tension. The diffusive (figure S3) and the stretching enhanced diffusion (figure S4) regimes are thus retrieved at small and large P\'eclet, respectively (figure \ref{fig:ligament_t12}a, circles).

If now fragmentation happens before the equilibration time predicted by equation \ref{eq:t12_hypo0_ligament}, we find that equilibration is controlled by the time required to develop the Rayleigh-Plateau capillary instability and break the ligament.
At low $\mathrm{We_1}$ and $\mathrm{Oh_{\kappa_1}}$, we indeed find that $t_{1/2}/t_{\dot\epsilon}\sim\mathrm{We_1}^{1/2}$ (figure \ref{fig:ligament_t12}b) and $t_{1/2}/t_{\kappa} \sim \mathrm{Oh_{\kappa_1}}$ (figure \ref{fig:ligament_t12}c), which correspond to $t_{1/2} \sim t_\gamma$.
This suggest that the flow associated with the fragmentation of the ligament into drops is strong enough to allow for fast equilibration of the drops.

\subsection{A free falling droplet}
\label{sec:basic-droplet}

At the end of the fragmentation sequence of the impactors' cores (figure \ref{fig:sed}), the fate of the remaining heat and chemical elements in the droplets depends on the P\'eclet number, \textit{i.e.} if they equilibrate in the high Pe spherical or in the diffusive regime.
In the diffusive regime (figure S6), the equilibration time scales as the typical diffusion time of the drop $t_{\kappa}=s_0^2/K_1$, irrespectively of viscosity and surface tension (figure \ref{fig:drop_t12}).

In the high Pe spherical regime (figure S7), equilibration is accelerated by the formation of thin thermal or compositinal boundary layers on the leading side of the drop, on both sides of the interface, where advection maintains a strong radial gradient of $\chi$. 
At a viscosity ratio $\eta_1/\eta_2$ on the order of 1 or smaller, equilibration is aided by the circulation forced within the drop by the viscous stress at the interface \citep{ulvrova_2011}.
The outer boundary layer is connected to a wake tail on the rear side of the drop.
A prediction for the equilibration time is obtained from mass conservation on the spherical drop and estimates of the convective flux (see appendix C for details). We find that
\begin{eqnarray}
t_{1/2} \sim \frac{s_0^2 q_1}{6 k_2}\mathrm{Pe_2}^{-1/2}\left[1+\left(\frac{q_2k_2}{q_1k_1}\right)^{1/2}\right] \text{regarding } \chi,\\
\label{eq:analytic_t12_drop_chi}
t_{1/2} \sim \frac{s_0^2 \rho_1c_{p_1}}{6 \lambda_2}\mathrm{Pe_2}^{-1/2}\left[1+\left(\frac{\rho_2c_{p_2}\lambda_2}{\rho_1c_{p_1}\lambda_1}\right)^{1/2}\right]  \text{regarding } T,\\
\label{eq:analytic_t12_drop_T}
t_{1/2} \sim \frac{s_0^2 D_{1/2}}{6 \kappa_2^c}\mathrm{Pe_2}^{-1/2}\left[1+\frac{1}{D_{1/2}}\left(\frac{\kappa_2^c}{\kappa_1^c}\right)^{1/2}\right]  \text{regarding } C.
\label{eq:analytic_t12_drop_C}
\end{eqnarray}
Equation \ref{eq:analytic_t12_drop_C} simplify as $t_{1/2} \sim (s_0^2 D_{1/2})/(6 \kappa_2^c)\mathrm{Pe_2}^{-1/2}$ and $t_{1/2} \sim (s_0^2/6)(\kappa_1^c\kappa_2^c)^{-1/2}\mathrm{Pe_2}^{-1/2}$ in the limits of siderophile ($D_{1/2}\gg 1$) and lithophile ($D_{1/2}\ll 1$) elements, respectively.
The equilibration time we predict for siderophile elements depends linearly on the partition coefficient, which is consistent with the numerical results of \citet{ulvrova_2011}. On the contrary, the equilibration time obtained for lithophile elements is independent of the partition coefficient, consistently with the analysis of \citet{samuel_2012}.

\section{Stirring Regime}
\label{sec:stirring}

\subsection{Deformation and stretching dynamic}
\label{sec:stirring_dynamic}

We now focus on the stirring regime identified in section \ref{sec:path}, which we think is the most relevant for metal segregation in a magma ocean \citep{dahl_2010,deguen_2011,deguen_2014,landeau_2014}. 

\begin{figure}[ht!]
\centering
\includegraphics[width=\linewidth]{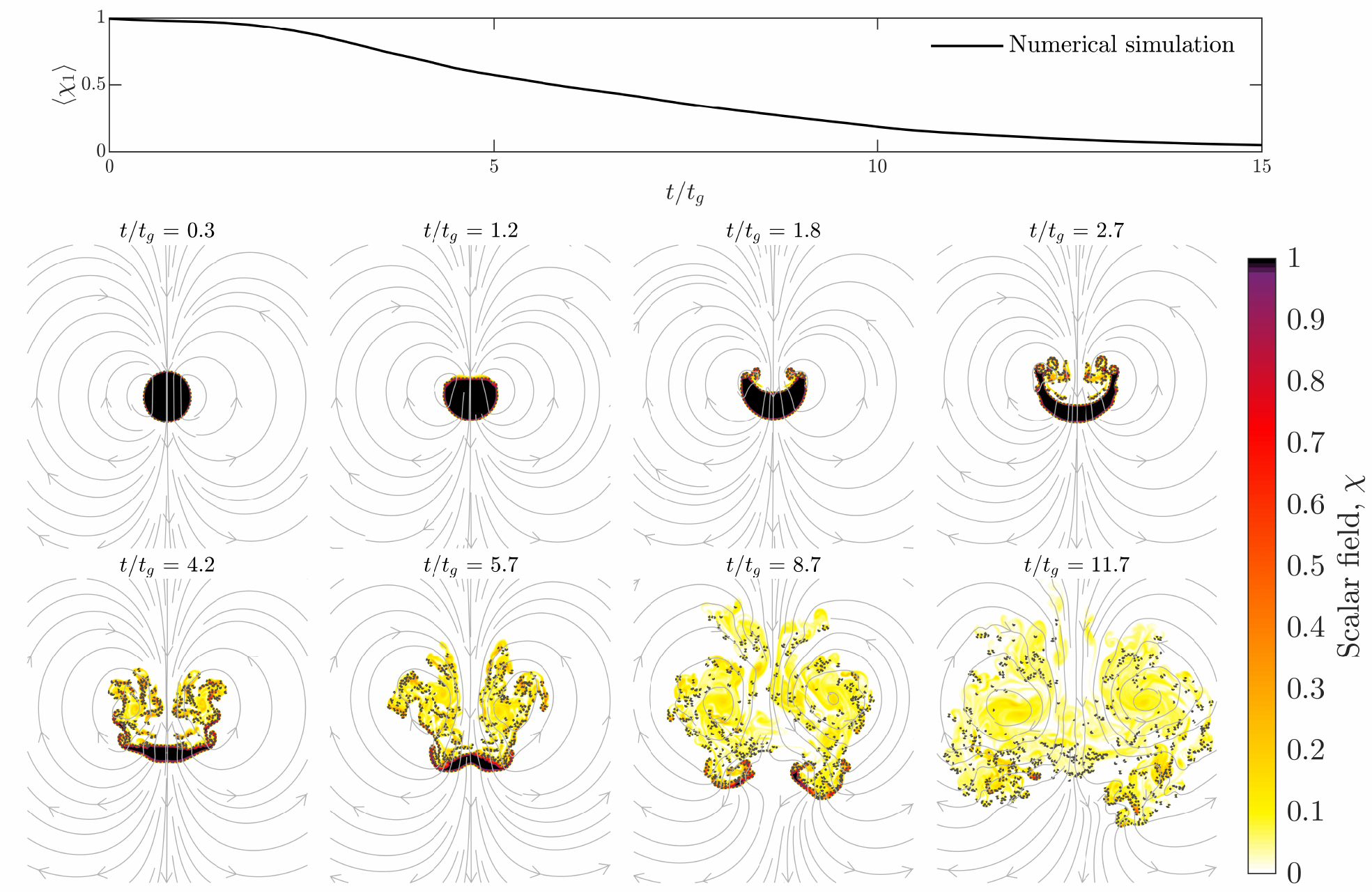} 
\caption{Stirring regime, $\mathrm{\mathrm{Pe_1}}=10^{4}$, $\mathrm{\mathrm{Re_1}}=10^{4}$ and $\mathrm{Bo}=10^3$. Upper: average value of $\chi_1$ in phase 1 as a function of time normalized by $t_{g}=\sqrt{(\rho_1/\Delta\rho)s_0/g}$. Lower: Snapshots showing maps of $\chi$ at different times. The dashed gray lines correspond to the interface position between phases 1 and 2. The gray lines correspond to the streamlines of the velocity field on a frame moving with the center of mass of the metal.}
\label{fig:drop_scalar_stirring}
\end{figure}

Figure \ref{fig:drop_scalar_stirring} shows snapshots of a simulation in this regime, at $\mathrm{\mathrm{Re_1}}=10^{4}$, $\mathrm{Bo}=10^3$, and $\mathrm{\mathrm{Pe_1}}=10^{4}$. 
The initially round drop quickly deforms, both because of an instability  which we interpret as a combination of Kelvin-Helmholtz (shear-driven) and Rayleigh-Taylor (buoyancy-driven) instabilities, and because of the action of the mean flow, which takes the form of a pair of expanding counter-rotating vortices (the 2D analogue of a vortex ring).
The interaction between the vortices and the interface instability leads to strong deformation and stretching of the metal phase, which topology evolves toward a collection of convoluted sheets (see for example figure \ref{fig:drop_scalar_stirring} at $t/t_g=4.2$ and figure \ref{fig:strain_map}b).
In our 2D numerical calculations, the metal phase eventually becomes discontinuous, but this ``fragmentation'' is an artifact due to the resolution limits of the computation grid. 2D sheets are stable against Rayleigh-Plateau capillary instabilities and are not expected to break.
In the simulations, the metal phase becomes discontinuous when the stretched structures become locally too thin to be resolved, but this is not a physical effect. A finer mesh indeed delays this apparent fragmentation. We therefore use the maximum resolution possible, but 2D simulations are fundamentally not designed to address fragmentation.
Nonetheless, these calculations are relevant for understanding the interplay between stretching and thermochemical equilibration, considering scales larger than the resolution limit. The Batchelor scale (\textit{e.g.} equation \ref{eq:s12_sheet}), estimated using a large scale stretching rate related to the pair of counter-rotating vortices, is indeed two order of magnitude larger than the resolution of the grid.

\subsection{Equilibration time}
\label{sec:stirring_equilibration}

The development of sheets from the initial volume of metal can be considered as the superposition of isolated stretched sheets with uniform stretching similar to that of section \ref{sec:basic-sheet}. We thus expect the equilibration time to be of the form $t_{1/2} \sim t_{\dot{\epsilon}}\mathrm{ln(Pe_1  \tau_{1/2})}$ (eq. \ref{eq:t12_hypo1}), where $\dot{\epsilon}$ is the stretching rate of the sheets, which needs to be estimated as a function of the flow geometry and strength. $\tau_{1/2}$ is a function of $p$ in 2D (section \ref{sec:basic-sheet} and figure \ref{fig:p_discussion}d), or a function of $K_1/K_2$ and $k_1/k_2$ in 3D (section \ref{sec:basic-ligament-analytical}), owing to the presence of both sheets and ligaments.
The flow has a vortex ring structure (figure \ref{fig:drop_scalar_stirring}), with a falling velocity $\langle u_y \rangle \sim \sqrt{(\Delta\rho /\rho_1) g s_0}$ varying on a scale $s_0$. 
If stretching is governed by the vortices, then $\dot{\epsilon}$ should be on the order of $\langle u_y \rangle/s_0$, which is equivalent to a stretching timescale equal to the free-fall timescale of the drop $t_g$.
This leads to
\begin{equation}
\label{eq:t12_drop_stirring}
t_{1/2} \sim t_g\mathrm{ln(Pe_1  \tau_{1/2})}.
\end{equation}
There is a good agreement between this prediction and the results from the numerical calculations of section \ref{sec:path}, which shows that in the stirring regime $t_{1/2}/t_g$ is independent of Re$_{1}$ (figure \ref{fig:drop_t12}b, diamonds) and Bo (figure \ref{fig:drop_t12}c, diamonds), and has a Pe$_{1}$ dependency (figure \ref{fig:drop_t12}a, dash-dotted line) well  fitted  by $t_{1/2}=0.53 t_g\mathrm{ln(Pe_1)}$.
This suggests that equilibration is controlled by stretching enhanced diffusion, with a stretching rate controlled by the large scale component of the flow. 

\subsection{The role of stretching on equilibration}
\label{sec:stirring_integral}

To address the role of stretching enhanced diffusion, we focus on the rate of gradient production \citep{ricard_2015}, obtained by applying the operator $2\nabla\chi\cdot\nabla$ to the advection-diffusion equation (eq. \ref{eq:AD}) and integrating over the 2D numerical domain $\Omega$ 
\begin{equation}
\label{eq:eq_ricard}
\frac{\mathrm{d}}{\mathrm{d}t}\int_\Omega|{\bf \nabla} \chi|^2\mathrm{d}S=-2\int_\Omega{\bf \nabla} \chi\cdot{\underline{\dot{\epsilon}}}\cdot{\bf \nabla} \chi\mathrm{d}S-2K\int_\Omega(\nabla^2\chi)^2\mathrm{d}S,
\end{equation}
where $\underline{\dot{\epsilon}}$ is the strain rate tensor.
The rate of gradient production (term on the left-hand side) is equal to the sum of an advection term (first term on the right-hand side) and a diffusion term (second term on the right-hand side). The advection term is related to the properties of the flow through the strain rate tensor $\underline{\dot{\epsilon}}$ and corresponds to the source of gradient production. It can be either positive or negative depending on the geometry of the flow and scalar field. In contrast, the diffusion term is always negative and damps scalar gradients.

\begin{figure}[ht!]
\centering
\includegraphics[width=1\linewidth]{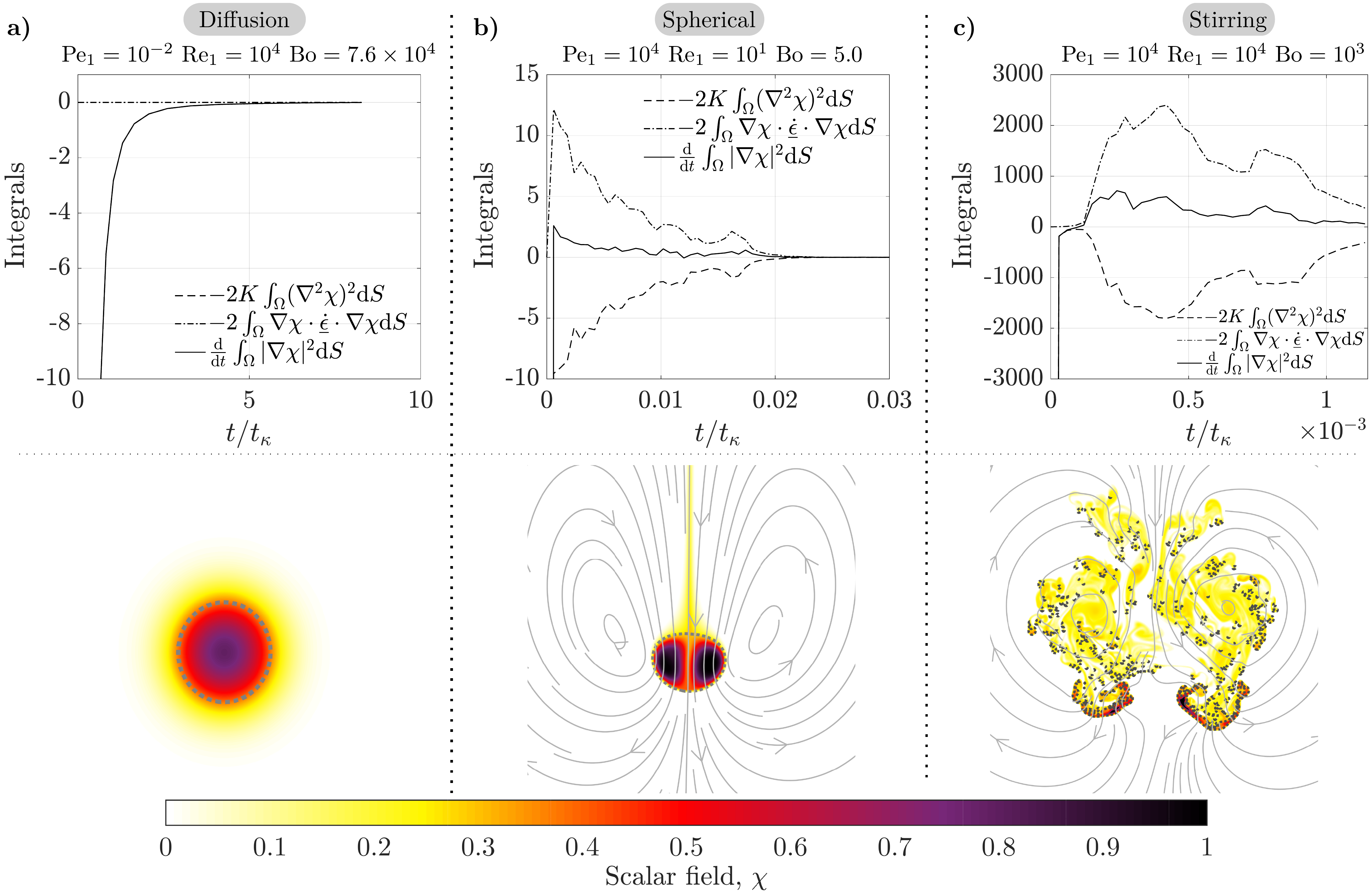}
\caption{Upper: Time evolution of the terms in equation \ref{eq:eq_ricard}, \textit{i.e.} the rate of gradient production, the advection term and the diffusion term \citep{ricard_2015}, in the small Pe diffusive (a), the high Pe spherical (b) and the stirring (c) regime. Lower: Snapshots showing maps of $\chi_i$ in these regimes. The dashed gray lines correspond to the interface position between phases 1 and 2. The gray lines correspond to the streamlines of the velocity field on a frame moving with the center of mass of the metal.}
\label{fig:integral}
\end{figure}

These integrals are calculated in three examples corresponding to small Pe diffusive, high Pe spherical and stirring regime:

(i) In the small Pe diffusive regime (figure \ref{fig:integral}a), the advection term is equal to zero and the rate of gradient production is equal to the diffusion term because scalar gradients are only dissipated and never generated. The diffusion term converges toward zero over time owing to the progressive equilibration of the drop, on a timescale $\sim t_{\kappa}$.

(ii) In the high Pe spherical regime (figure \ref{fig:integral}b), the advection term is positive because scalar gradients are produced by advection in the boundary layer in front of the drop during its fall. 
Starting from 0, the advection term first increases during a short transient during which the thermal boundary layer grows by diffusion up to a thickness at which advection balances diffusion. 
The advection term then slowly decrease due to the decrease of difference of $\chi$ between the drop and the surrounding, while the thickness of the thermal boundary layer remains constant.
The magnitude of the diffusion term follows closely the evolution of the advection term, but is slightly smaller, which shows that  equilibration is controlled by the advection  and gradients production. 
The advection term and the diffusion term converge toward zero as the drop equilibrates with the surrounding fluid, with an equilibration time about 100 times smaller than the diffusion time.
In the simulation, the shape of the drop slightly varies in time, explaining the fluctuations around the mean exponential decrease obtained on the integrals.

(iii) In the stirring regime (figure \ref{fig:integral}c), the advection term becomes positive owing to the production of scalar gradients by the stretching. The advection term is larger than the diffusion term as soon as deformation and stretching of the drop develop, leading to a positive rate of gradient production. Equilibration is then controlled by the production of scalar gradients (whose stirring is the source). At first, the rate of gradient production increases, in connection with the development of stretched structures through hydrodynamics instabilities and stirring. Then, the rate of gradient production decreases owing to the diffusive equilibration of the metal phase with the surrounding fluid when the thickness of the sheets become small enough.

\begin{figure}[ht!]
\centering
\includegraphics[width=1\linewidth]{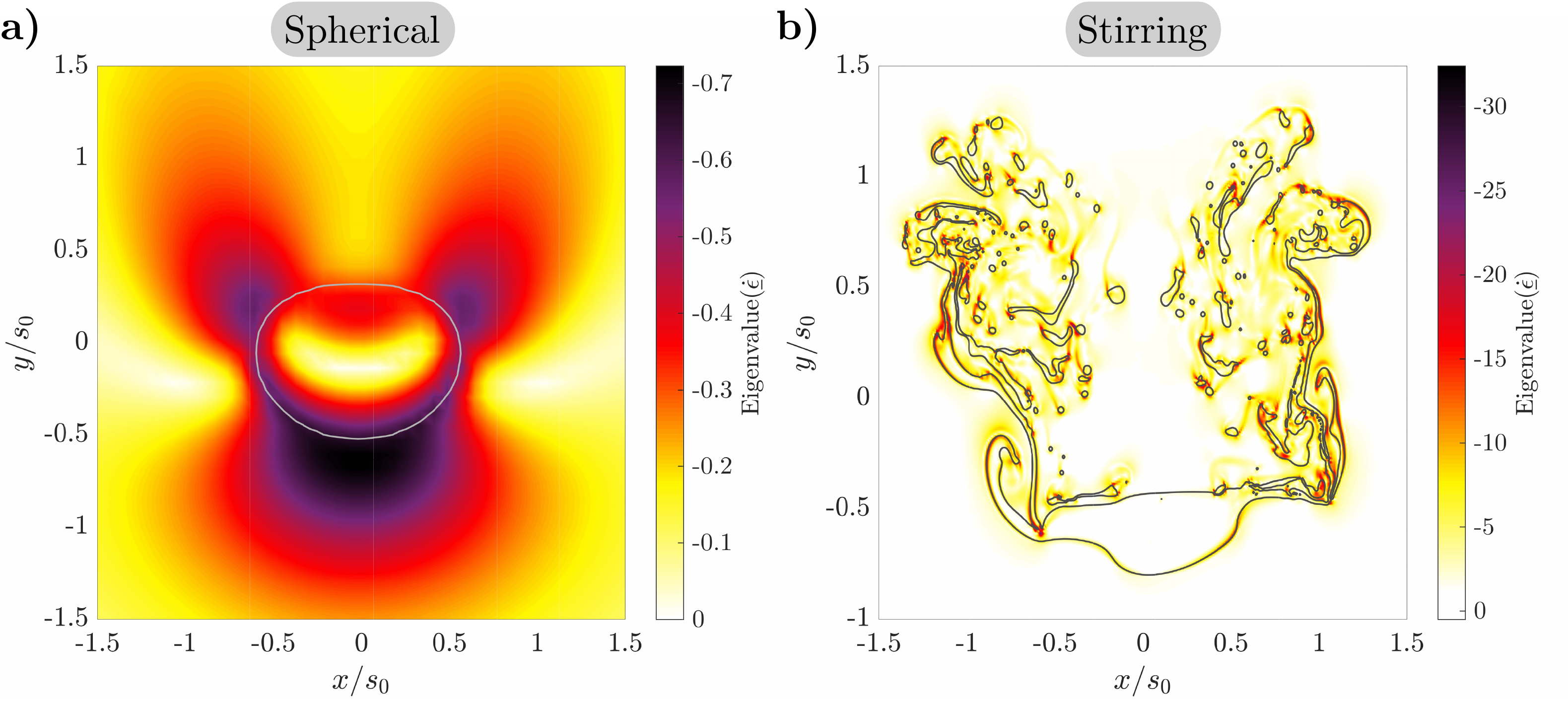}
\caption{Map of the local negative eigenvalue of the strain rate tensor $\underline{\dot{\epsilon}}$ in the high Pe spherical (a) and the stirring regime (b). The gray lines corresponds to the interface between the metallic and the silicate phase.}
\label{fig:strain_map}
\end{figure}

Figure \ref{fig:strain_map} shows the negative eigenvalue of the strain rate tensor $\underline{\dot{\epsilon}}$, which we interpret as a stretching magnitude \citep{ricard_2015}.
In the high Pe spherical regime (a), stretching develops on a scale comparable with the size of the drop, which is significantly larger than the thin boundary layer where thermal and compositional exchanges occur.
On the contrary, in the stirring regime (b), stretching occurs in the vicinity of the interface, over a very localized deformation area. The source of gradient production responsible for the large advection term (figure \ref{fig:integral}c, upper) is then located at the same place as the interface deformation. Small-scale stretching processes related to the stirring are responsible for gradient production, arguing in turn for an equilibration controlled by stretching enhanced diffusion.

\subsection{Probability Distribution Functions of $\chi$}
\label{sec:stirring_pdf}

\begin{figure}[ht!]
\centering
\includegraphics[width=1\linewidth]{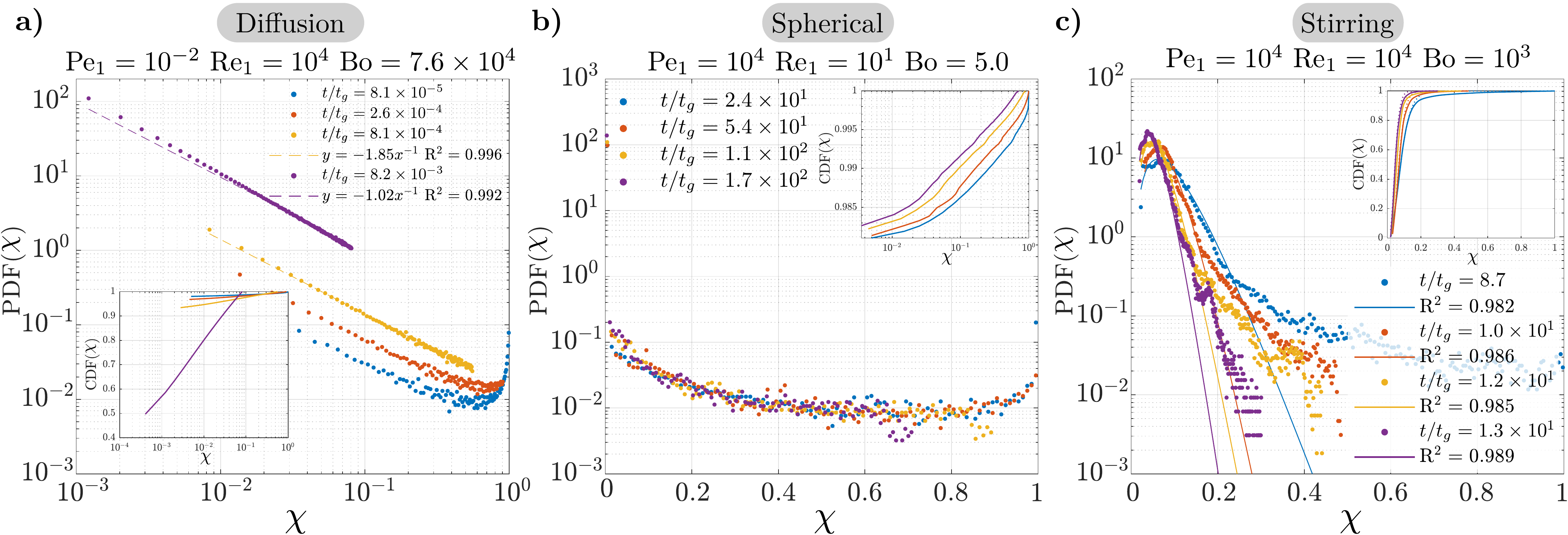} 
\caption{Time evolution of the probability distribution function (PDF) of the scalar field in the small Pe diffusive (a), the high Pe spherical (b) and the stirring regime (c). Insets correspond to the time evolution of the cumulative distribution function (CDF) of the scalar field. The solid lines correspond to the fitted one parameter gamma distributions (eq. \ref{eq:gamma_distrib}).}
\label{fig:stats}
\end{figure}

The qualitative difference between the three regimes is also highlighted by the probability distribution functions (PDF) of the scalar field.
In the small Pe diffusive regime (figure \ref{fig:stats}a), the distribution of $\chi$ tends toward a probability density $\propto 1/\chi$.
This is consistent with the radial dependency of $\chi$, which, after the initially discontinuous radial profile of  $\chi$ has smoothed out, tends toward a self-similar profile of the form $\chi(r,t)\propto \exp(-r^{2}/(K_{2} t))$. 
It can be verified that the PDF of a field with this spatial distribution is indeed $\propto 1/\chi$.
In the high Pe spherical regime (figure \ref{fig:stats}b), the PDF slowly shifts toward smaller scalar values over time, conserving its shape consistently with the stable evolution of the scalar field inside the drop during its fall.
In the stirring regime (figure \ref{fig:stats}c), $\chi$ has a peaked distribution, which over time becomes narrower and shifts toward lower scalar levels as a result of the progressive homogenization.

It has been argued \citep{duplat_2010,meunier_2003,villermaux_2004} that the interaction between stretched  sheets and ligaments lead to the addition of their scalar fields and to the stable self-convolution of their scalar distributions.
It is predicted that in this regime the scalar field distribution follows a one parameter gamma distribution,
\begin{equation}
P(X=\chi/\langle \chi\rangle)=\frac{n^n}{\Gamma(n)}X^{n-1}e^{-nX},
\label{eq:gamma_distrib}
\end{equation}
where $n=1/\sigma^2$ is related to the standard deviation $\sigma$ of the distribution.
After a time corresponding to the development of stretched structures, the PDFs from our calculations are indeed well approximated by these one parameters gamma distributions (figure \ref{fig:stats}c), 
except in the tail where the probability density is higher. 
This again is consistent with an equilibration controlled by stretching enhanced diffusion. 
Self-convolution as a result of scalar fields addition also supports the assumption on the superposition of isolated sheets we made to predict the equilibration time of the drop (eq. \ref{eq:t12_drop_stirring}).

\section{Conclusion}
\label{sec:conclusion}

Deformation and fragmentation of differentiated impactors' cores are key mechanisms for heat and chemical elements transfer during the differentiation of planetary bodies. 
We argue that fragmentation of the metallic core of an impactor must involves a change of topology leading to a collection of sheets and ligaments, which can then fragment into drops (figure \ref{fig:sed}).
This picture is consistent with previous experimental results \citep{deguen_2011,deguen_2014,kendall_2016,landeau_2014,wacheul_2014,wacheul_2018} and our numerical simulations (section \ref{sec:stirring_dynamic}), which suggest that a large volume of metal falling into a magma ocean would be vigorously stirred and stretched. 
We focus here on the physical concepts associated with the fragmentation and equilibration sequence. The ``building blocks'' of the fragmentation sequence involve stretched structures (sheets and ligaments) in which stretching enhanced diffusion, generalized here to immiscible fluids, improves significantly chemical and thermal equilibration efficiency between metal and silicates. The equilibration time depends mainly on the stretching rate with, if the stretching rate is constant, a logarithmic dependency on diffusivity and partition coefficient.
We find that thermal equilibration may occur before fragmentation whereas chemical equilibration may still be incomplete at this moment.
Fragmentation eventually produces droplets whose chemical equilibration depends critically on the partition coefficient (equation \ref{eq:analytic_t12_drop_C}).

In the stirring regime identified for large volume of metals (in the sense of having large Reynolds and Bond numbers), the metal phase is vigorously stretched and deformed. We find from numerical simulations in that regime that the equilibration time depends mainly on the stretching rate of the metal phase and weakly on diffusivity, irrespectively of viscosity and surface tension.
Our results suggest that thermochemical equilibration is controlled by the stretching enhanced diffusion process, which in our numerical simulations develops from the mean flow, as the fall of the drop entrains surrounding silicates.
We find that in our numerical calculations the equilibration time is well approximated by $t_{1/2} \sim \sqrt{\frac{\rho_{1}}{\Delta\rho}\frac{s_{0}}{g}} \mathrm{ln\, Pe_1}$ (equation  \ref{eq:t12_drop_stirring}), where the P\'eclet number is built on a stretching rate equal to $\langle u_y \rangle/s_0=\sqrt{(\Delta\rho/\rho_1)gs_0}/s_0$.
Using this scaling would predict relatively large equilibration distances (50\% of the initial temperature/composition difference) in a magma ocean:  
as an example,  the composition and temperature equilibration distances of a 1 km impactor's core sinking in a magma ocean would be 12 km and 8 km, respectively.
Equation \ref{eq:t12_drop_stirring} could also be used to predict that only impactors with a core smaller than 56 km and 41 km will significantly equilibrate thermally or chemically in a 3000 km deep magma ocean.

However, these predictions are based on the assumption  that the stretching rate is governed by the mean flow (eq. \ref{eq:t12_drop_stirring}).
This seems to be correct in our 2D calculations, but is unlikely to hold in 3D at higher Re and Bo due to the development of turbulent fluctuations. Turbulent velocity fluctuations would increase the stretching of the metal phase, and thus decrease the equilibration time. 
The prediction of equation \ref{eq:t12_drop_stirring} therefore most probably overestimates significantly the equilibration time. 
We may anticipate that in the limit of high Re$_{1}$, Bo and Pe$_{1}$, equilibration of the metal with the surrounding silicates would even be fast enough for the evolution of the mean value of $\chi$ in the metal phase to be limited by the rate of entrainement of fresh silicates rather than by equilibration at the local scale.
Our analysis suggests that the key for testing this hypothesis and estimating the equilibration times is to obtain models predicting the stretching rate of the metal phase during the cratering phase and post-cratering flow.

\appendix
\section{Solution of the Diffusion Equation for the Stretched Sheet}
\label{app:sheet}

The solution of the diffusion equation \ref{eq:D1D} for a stretched isolated sheet is given in each phase by
\begin{eqnarray}
\nonumber\chi_1(\xi\leq 1/2,\tau)&=&\frac{\Delta\chi}{2}\Bigg\lbrace-(1+p)~\mathrm{erf}\left(\frac{\xi-1/2}{2\sqrt{\tau}}\right)+(1+p)\\
&&\cdot\sum_{n=1}^\infty (-p)^{n-1}\bigg[\mathrm{erf}\left(\frac{n+\xi-1/2}{2\sqrt{\tau}}\right)-p~\mathrm{erf}\left(\frac{n-\xi+1/2}{2\sqrt{\tau}}\right)\bigg]\Bigg\rbrace,\\
\label{eq:analytic_full_1}
\nonumber\chi_2(\xi\geq 1/2,\tau)&=&\frac{\Delta\chi}{2}\Bigg\lbrace-(1-p)~\mathrm{erf}\left(\frac{\xi-1/2}{2\sqrt{K_2/K_1\tau}}\right)+(1-p^2)\\
&&\cdot\sum_{n=1}^\infty (-p)^{n-1}~\mathrm{erf}\left(\frac{\sqrt{K_2/K_1}n+\xi-1/2}{2\sqrt{K_2/K_1\tau}}\right)\Bigg\rbrace.
\label{eq:analytic_full_2}
\end{eqnarray}

\section{Solution of the Diffusion Equation for the Stretched Ligament}
\label{app:ligament}

The solution of the diffusion equation for a stretched isolated ligament is given in each phase by
\begin{eqnarray}
\label{eq:analytic_cylinder1}
\chi_1(\xi \leq 1,\tau)&=&\frac{4\Delta\chi k^2}{\pi^2 K^2}\int_0^\infty \mathrm{e}^{-u^2\tau}\frac{J_0(u\xi)J_1(u)\mathrm{d}u}{u^2[\phi^2(u)+\psi^2(u)]},\\
\label{eq:analytic_cylinder2}
\chi_2(\xi \geq 1,\tau)&=&\frac{2\Delta\chi k^2}{\pi K}\int_0^\infty \mathrm{e}^{-u^2\tau}J_1(u)\frac{J_0(K u\xi)\phi(u)-Y_0(K u\xi)\psi(u)}{u[\phi^2(u)+\psi^2(u)]}\mathrm{d}u,
\end{eqnarray}
where $K=\sqrt{K_1/K_2}$, $k=\sqrt{k_1/k_2}$ and
\begin{eqnarray}
\label{eq:analytic_cylinder3}
\psi(u)&=&\frac{k^2}{K}J_1(u)J_0(Ku)-J_0(u)J_1(Ku),\\
\label{eq:analytic_cylinder4}
\phi(u)&=&\frac{k^2}{K}J_1(u)Y_0(Ku)-J_0(u)Y_1(Ku).
\end{eqnarray}
$J_\nu$ and $Y_\nu$ are respectively the Bessel function of the first and the second kind, with $\nu$ their order.

\section{Equilibration Time for the Free-Falling Droplet}
\label{app:droplet}

We consider here the equilibration of a drop of diameter $s_0$ and falling velocity $u$, with internal circulation, at high Pe$_{1}$ and Pe$_{2}$.
In this limit, we expect the formation of two thin 
thermal/compositional boundary layers, on both sides of the droplet interface, with respective thicknesses $\delta_1$ and $\delta_2$. 
Depending on the properties of the two phases, the flux can be limited by one or the other of the boundaries.
Denoting by $\left< \chi_1 \right>$ the mean of $\chi_{1}$ in the drop, by $\chi_{i}$ the value of $\chi$ at the interface, and by $\chi_2^{\infty}$ the value of $\chi_{2}$ far from the drop, the heat/mass flux $\Phi$ across the interface is 
\begin{equation}
\label{eq:Phi}
    \Phi \sim k_1\frac{\left< \chi_1 \right>-\chi_i}{\delta_1} \sim k_2\frac{\chi_i-\chi_{2}^{\infty}}{\delta_2}.
\end{equation}
The boundary layers thickness on both sides of the interface can be shown to be $\delta_i \sim \sqrt{K_is_0/u} \sim s_0\, \mathrm{Pe}_i^{-1/2}$ by balancing advection and diffusion in the vicinity of the interface (see for example \citet{ribe_2015} for the outer boundary layer scaling; a similar reasoning leads to the scaling of the inner boundary layer, which has the same form).
Writing $\chi_i$ as a function of $\left< \chi_1 \right>$ and $\chi_2^{\infty}$ from equation \ref{eq:Phi},  and noting that $\delta_1/\delta_2 \sim \sqrt{K_1/K_2}$, the flux becomes
\begin{equation}
    \Phi \sim \frac{k_2}{s_0}\mathrm{Pe}_2^{1/2}\frac{\left< \chi_1 \right>-\chi_2^{\infty}}{1+\left(\frac{q_2k_2}{q_1k_1}\right)^{1/2}},
\end{equation}
which is consistent with the heat flux given by \citet{ribe_2015} in the limit $q_1 k_1 \gg q_2 k_2$. Using conservation of $\chi$ gives
\begin{equation}
    \frac{\pi}{6} s_0^3 q_1\frac{\mathrm{d} \left< \chi_1 \right>}{\mathrm{d}t}=-\pi s_0^2\Phi,
\end{equation}
from which we obtain the equilibration time
\begin{equation}
    t_{1/2} \sim \frac{s_0^2 q_1}{6k_2}\mathrm{Pe_2}^{-1/2}\left[1+\left(\frac{q_2k_2}{q_1k_1}\right)^{1/2}\right].
\end{equation}

\acknowledgments
This project has received funding from the European Research Council (ERC) under the European Unions Horizon 2020 research and innovation programme (grant agreement No 716429). T. Alboussi\`ere provided comments that improved the manuscript. We thank Jean-Baptiste Wacheul and an anonymous reviewer for their useful comments. The codes are available at https://figshare.com/articles/lherm\_basilisk\_stretching/6850271 and basilisk.fr.

%
%
%
%
%
%
%
%
%






\end{document}